\documentclass[10pt]{article}
\usepackage{amsmath}
\usepackage{graphicx}
\usepackage{amsfonts}
\usepackage{amssymb}
\usepackage{epsf}
\usepackage{latexsym}

\def\d{\mbox{d}}

\def\sc{{\mbox{\scriptsize c}}}

\def\ma{\mbox{a}}
\def\mb{\mbox{b}}

\def\md{\mbox{d}}

\def\mj{\mbox{j}}

\def\mo{\mbox{o}}
 
\def\mq{\mbox{q}}

\def\mt{\mbox{t}}

\def\mC{\mbox{C}}

\def\mF{\mbox{F}}

\def\mH{\mbox{H}}

\def\mP{\mbox{P}}

\def\mR{\mbox{R}}
\def\mS{\mbox{S}}

\def\be{\begin{equation}}
\def\beq{\begin{equation}}
\def\ee{\end{equation}}
\def\eeq{\end{equation}}
\def\bea{\begin{eqnarray}}
\def\eea{\end{eqnarray}}

\def\pa{\partial}
\def\d{\textrm{d}}

\def\cr{\mbox{\scriptsize{\bf $\mbox{ } \times \mbox{ }$}}}

\def\hat{\widehat}
\def\brho{\mbox{\boldmath$\rho$}}

\def\sa{\mbox{\scriptsize a}}

\def\sc{\mbox{\scriptsize c}}
\def\sd{\mbox{\scriptsize d}}
\def\sa{\mbox{\scriptsize a}}

\def\sc{\mbox{\scriptsize c}}
\def\sd{\mbox{\scriptsize d}}
 
\def\sh{\mbox{\scriptsize h}} 
\def\sj{\mbox{\scriptsize j}} 
\def\sk{\mbox{\scriptsize k}}
\def\sll{\mbox{\scriptsize l}} 
\def\sss{\mbox{\scriptsize s}}  
\def\se{\mbox{\scriptsize e}}

\def\sm{\mbox{\scriptsize m}}
\def\sn{\mbox{\scriptsize n}} 
\def\so{\mbox{\scriptsize o}} 
\def\sp{\mbox{\scriptsize p}}

\def\sr{\mbox{\scriptsize r}}
\def\ss{\mbox{\scriptsize s}}
\def\st{\mbox{\scriptsize t}}
\def\su{\mbox{\scriptsize u}}

\def\sE{\mbox{\scriptsize E}}
\def\sF{\mbox{\scriptsize F}}
\def\sG{\mbox{\scriptsize G}}
\def\sH{\mbox{\scriptsize H}}

\def\sR{\mbox{\scriptsize R}}
\def\sS{\mbox{\scriptsize S}}

\def\sY{\mbox{\scriptsize Y}} 
 
\def\eph{\mbox{\scriptsize eph}}
 
\def\eph(B){\mbox{\scriptsize emergent(LMB)}}

\def\fD{\mbox{\sffamily D}}
\def\fE{\mbox{\sffamily E}}

\def\fH{\mbox{\sffamily H}}

\def\fJ{\mbox{\sffamily J}}

\def\fQ{\mbox{\sffamily Q}}
\def\fR{\mbox{\sffamily R}}
\def\fS{\mbox{\sffamily S}}
\def\fT{\mbox{\sffamily T}}
\def\fU{\mbox{\sffamily U}}
\def\fV{\mbox{\sffamily V}}
\def\fW{\mbox{\sffamily W}}

\def\sfA{\mbox{\sffamily{\scriptsize A}}}

\def\sfS{\mbox{\sffamily{\scriptsize S}}}

\begin{document}

\begin{titlepage}

\begin{center}

\vspace{.3in}

{\large{\bf ON THE SEMICLASSICAL APPROACH TO QUANTUM COSMOLOGY}} 

\vspace{.3in}

{{\bf Edward Anderson$^*$}}

\vspace{.3in}

{\sl $^1$ Departamento de F\'{\i}sica Te\'{o}rica, Universidad Autonoma de Madrid.}

\end{center}

\vspace{.3in}

%\baselineskip=24pt 

%===========================================================ABSTRACT=======================================================================
\begin{abstract}

\noindent The emergent semiclassical time approach to resolving the problem of time in quantum gravity 
involves heavy slow degrees of freedom providing via an approximately Hamilton--Jacobi equation an 
approximate timestandard with respect to which the quantum mechanics of light fast degrees of freedom 
can run.  
More concretely, this approach involves Born--Oppenheimer and WKB ans\"{a}tze and some accompanying approximations.  
In this paper, I investigate this approach for concrete scaled relational particle mechanics models, i.e. 
models featuring only relative separations, relative angles and relative times.  
I consider the heavy--light interaction term in the light quantum equation -- necessary for the semiclassical 
approach to work, firstly as an emergent-time dependent perturbation of the emergent-time-dependent Schr\"{o}dinger 
equation for the light subsystem. 
Secondly, I consider a scheme in which the backreaction is small but non-negligible, so that the $l$-subsystem 
also affects the form of the emergent time.  
I also suggest that the many terms involving expectation values of the light wavefunctions in both the 
(unapproximated) heavy and light equations might require treatment in parallel to the Hartree--Fock self-consistent 
approach rather than merely being discarded; for the moment this paper provides a counterexample to such terms 
being smaller than their unaveraged counterparts.  
Investigation of these ideas and methods will give us a more robust understanding of the suggested
quantum-cosmological origin of microwave background inhomogeneities and galaxies.  

\end{abstract}

%==========================================================================================================================================

\mbox{ }

%===============================================COMMENTS FOR PREPRINT ARCHIVE==============================================================

%\noindent{\bf Keywords:} 

%==========================================================================================================================================

\noindent{\bf PACS numbers 04.60-m, 04.60.Ds, 98.80.Qc}

\vspace{2in}

\noindent$^*$ ea212@cam.ac.uk, edward.anderson@uam.es

\end{titlepage}

%========================================================================================================
%==========================================================================================================================================
\section{Introduction}
%==========================================================================================================================================
%==========================================================================================================================================

This paper concerns the semiclassical approach to the Problem of Time in Quantum Gravity (see Sec 1.1) 
and to other conceptual issues in Quantum Cosmology (see Sec 1.2). 
I concentrate on a relational particle mechanics (RPM) model (what these are and the motivation for them 
is considered in Sec 1.3; how this is of use in the semiclassical approach is explained in Sec 1.4).

%========================================================================================================
\subsection{The Problem of Time in Quantum Gravity} 
%========================================================================================================

The Problem of Time \cite{Wheeler, K81, K92, I93, POTOther, Kieferbook, Rovellibook, APOT} follows from how what one means by `time' has a different meaning in each of General Relativity (GR) and ordinary Quantum Theory (QT).  
This incompatibility creates serious problems with trying to replace these two branches of physics with a single framework in regimes in which neither QT nor GR can be neglected, such as in black holes or in the very early universe.  
One well-known facet of the Problem of Time appears in attempting canonical quantization of GR due to the GR 
Hamiltonian constraint\footnote{$h_{\alpha\beta}$ is the spatial 3-metric,
%%%%%%%%%%%%%%%%%%%%%%%%%%%%%%%%%%%%%%%%%%%%%%%%%%%%%%%%%%%%%%%%%%%%%%%%%%%%%%%%%%%%%%%%%%%%%%%%%%%%%%%%%%%%%%%%%%%%% 
with determinant $h$, covariant derivative $D_{\alpha}$, Ricci scalar $R$ and conjugate momentum $\pi^{\alpha\beta}$.
${\cal M}^{\alpha\beta\gamma\delta} = \sqrt{h}\{h^{\alpha\gamma}h^{\beta\delta} - h^{\alpha\beta}h^{\gamma\delta}\}$ 
is the GR kinetic metric on the GR configuration space with determinant ${\cal M}$.
Its inverse 
${\cal N}_{\alpha\beta\gamma\delta} =\{h_{\alpha\gamma}h_{\beta\delta} - h_{\alpha\beta}h_{\gamma\delta}/2\}/\sqrt{h}$ 
is the DeWitt supermetric of GR.}
%%%%%%%%%%%%%%%%%%%%%%%%%%%%%%%%%%%%%%%%%%%%%%%%%%%%%%%%%%%%%%%%%%%%%%%%%%%%%%%%%%%%%%%%%%%%%%%%%%%%%%%%%%%%%%%%%%%%% 
\beq
{\cal H} := 
N_{\alpha\beta\gamma\delta}\pi^{\alpha\beta}\pi^{\gamma\delta} - \sqrt{h}R = 0  
\label{GRham}
\eeq 
being quadratic but not linear in the momenta. 
Then elevating ${\cal H}$ to a quantum equation produces a stationary, i.e. timeless or frozen wave equation: the Wheeler--DeWitt \cite{Wheeler, DeWitt} equation 
\beq
\hat{\cal H}\Psi = 
-\left\{
\frac{1}{\sqrt{{\cal M}}}\frac{\delta }{\delta h_{\mu\nu}}
\left\{
\sqrt{{\cal M}}{\cal N}_{\mu\nu\rho\sigma} 
\frac{\delta }{\delta h_{\rho\sigma}}
\right\} 
+ \sqrt{h}R  
\right\}\Psi
= 0 \mbox{ } , 
\label{WDE}
\eeq 
where $\Psi$ is the wavefunction of the Universe.  
Note that one gets this frozen equation of type $\widehat{\fH}\Psi = 0$ {\sl in place of} ordinary QT's time-dependent 
Schr\"{o}dinger equation, 
\beq
i\hbar\pa\Psi/\pa t = \hat{\fH}\Psi \mbox{ } .
\eeq 
(Here, I use $\fH$ to denote Hamiltonians, and $t$ for the absolute Newtonian time.) 
The above is, moreover, but one among various facets of the Problem of Time; see \cite{K92, I93, APOT} for discussion of others.

Many strategies have been tried, but none work when examined in detail.
Many of the technical difficulties and some of the conceptual difficulties (see e.g. 
\cite{APOT} for more) come from GR also possessing the momentum constraint, 
\beq
{\cal L}_{\mu} := -2D_{\nu}\pi^{\nu}\mbox{}_{\mu} = 0
\label{GRmom} \mbox{ } .  
\eeq

Some of the strategies toward resolving the Problem of Time are as follows.  

\mbox{ } 

\noindent A) It may still be that a classical time exists but happens to be harder to find.
We now consider starting one's scheme off by finding a way of solving ${\cal H}$ in general at the classical level 
(`tempus ante quantum') to obtain a part-linear form, schematically (ignoring complications from the momentum 
constraint),   
\beq
p_{t^{\sa\sn\st\se}} + 
\fH^{\st\sr\su\se}(x; t^{\sa\sn\st\se}(x), q_{\so\st\sh\se\sr}^{\Gamma}(x), 
p^{\so\st\sh\se\sr}_{\Gamma}(x)] = 0 
\mbox{ } ,
\label{PLIN}
\eeq
where $p_{t^{\sa\sn\st\se}}$ is the momentum conjugate to a candidate classical time variable, 
$t^{\sa\sn\st\se}$ which is to play a role parallel to that of external classical time.
$\fH^{\st\sr\su\se}$ is then the `true Hamiltonian' for the system.  
Given such a parabolic form for ${\cal H}$, it becomes possible to apply a conceptually-standard 
quantization that yields the time-dependent Schr\"{o}dinger equation 
\beq
i\pa\Psi/\pa t^{\sa\sn\st\se} = \hat{\fH}_{\st\sr\su\se}(x; t^{\sa\sn\st\se}(x), 
q_{\so\st\sh\se\sr}^{\Gamma}(x), p^{\so\st\sh\se\sr}_{\Gamma}(x)]\Psi \mbox{ } , 
\label{TEEDEE}
\eeq 
with obvious associated Schr\"{o}dinger inner product.
A first such suggestion is that there may be a {\it hidden} alias {\it internal} time \cite{York72, K81, 
K92, I93} within one's gravitational theory itself.  
It is to be found by applying some canonical transformation.
An example of such is York time.
There are also non-geometrodynamical internal time candidates: {\it Matter Time Approaches} (See e.g.  
\cite{K92, RefFlu}).  
I.e., one can consider extending the set of variables from the geometrodynamical ones to include 
also matter variables coupled to these, which then serve to label spacetime events.  

\mbox{ }

\noindent B) Perhaps there is no time in general at the classical level, but some notion of time 
emerges under certain circumstances at the quantum level.
E.g. slow, heavy `$h$'  variables can provide an approximate timefunction with respect to which the 
other fast, light `$l$' degrees of freedom evolve: {\it semiclassical approach} \cite{DeWitt, SemiclOther, 
HallHaw, BV89Kiefer94, K92, I93, Kiefer99, Kieferbook, SemiclI}.

As the semiclassical approach B) is the main focus of this paper, I explain this approach in further 
detail than the other ones.  
In Quantum Cosmology the role of $h$ is played by scale (and homogeneous matter modes).   
In the Halliwell--Hawking approach \cite{HallHaw} to Quantum Cosmology, the $l$-part are small 
inhomogeneities.  
This approach goes via making the Born--Oppenheimer ansatz 
\beq
\Psi(h, l) = \psi(h)|\chi(h, l)\rangle
\eeq 
and the WKB ansatz 
\beq
\psi(h) = \mbox{exp}(i\fW(h)/\hbar) \mbox{ } 
\eeq 
(each of which furthermore suggests a number of approximations).  
One then forms the $h$-equation 
\beq
\langle\chi| \hat{\cal H} \Psi = 0 \mbox{ } , 
\eeq 
which, under a number of simplifications, yields a Hamilton--Jacobi\footnote{For simplicity, this is
%%%%%%%%%%%%%%%%%%%%%%%%%%%%%%%%%%%%%%%%%%%%%%%%%%%%%%%%%%%%%%%%%%%%%%%%%%%%%%%%%%%%%%%%%%%%%%%%%%%%%%%%% 
presented in the case of one $h$ degree of freedom and with no linear constraints.} 
%%%%%%%%%%%%%%%%%%%%%%%%%%%%%%%%%%%%%%%%%%%%%%%%%%%%%%%%%%%%%%%%%%%%%%%%%%%%%%%%%%%%%%%%%%%%%%%%%%%%%%%%%
equation, i.e. an equation paralleling the equation 
\beq
\{{\pa \fW}/{\pa h}\}^2 = 2\{\fE - \fV(h)\} \mbox{ }  
\label{h-HJE}
\eeq
which is familiar from mechanics.
Here, $\fW$ is the characteristic function.   
Next, one approach to such a Hamilton--Jacobi-type equation is to solve it for an approximate emergent 
semiclassical time $t^{\se\sm} = t^{\se\sm}(h)$. 
Then the $l$-equation 
\beq
\{1 - |\chi\rangle\langle\chi|\}\hat{\cal H}\Psi = 0
\eeq 
can be recast (modulo further simplifications) as a $t^{\se\sm}$-dependent Schr\"{o}dinger equation for 
the $l$ degrees of freedom
\beq
i\hbar\pa|\chi\rangle/\pa t^{\se\sm}  = \widehat{\fH}_{l}|\chi\rangle \mbox{ }
\label{TDSE2} \mbox{ }  
\eeq
for $\fH_l$ the Hamiltonian for the $l$-subsystem.  
For detail of how this recasting works, the current paper's toy model case in Sec 3 suffices to give an understanding,  
so I refer the reader to there.

The `paradox' between the theoretical timelessness of the universe and the quotidian semblance of 
dynamics is discussed in e.g. \cite{Zeh, BS, BV89Kiefer94, K92, I93, B93, B94II, EOT, H99, H03, HT, 
SemiclI}.   
In particular, the semiclassical approach does not work out (e.g. \cite{SemiclI} reviews this) if one 
considers a more general `superposed' rather than WKB wavefunction ansatz, nor is there an entirely 
well-established a priori reason for the WKB wavefunction ansatz.  
Investigating this is further complicated by the `many approximations problem' \cite{SemiclI} -- there 
are of the order of 20 to 30 such needed in semiclassical schemes
There are a number of arguments about the importance of checking how backreaction terms, i.e. $l$-wavefunction 
dependent terms in the $h$-equation by which the $l$-subsystem influences back the $h$-system and the 
emergent time that approximately arises from it.  
E.g. in wishing to model GR, as a conceptually important part of the theory (in its aspect as 
supplanter of absolute structure) is for the matter to backreact on the geometry.  
There are also theory-independent reasons for backreaction terms -- how could one subsystem genuinely 
provide time for another if the two are not coupled to each other?  
The literature on the semiclassical approach makes common use of the ``neglecting averages" approximation.  
Moreover, dropping averaged terms turns out to substantially distort the outcome in Molecular Physics 
calculations (the Hartree--Fock self-consistent scheme).  

\mbox{ }

\noindent C) There are also a number of approaches that take timelessness at face value. 
Here, one considers only questions about the universe `being', rather than `becoming', a certain way.  
This can cause some practical limitations, but can address at least some questions of interest. 
E.g., the {\it na\"{\i}ve Schr\"{o}dinger interpretation} \cite{HP86UW89} concerns the `being' 
probabilities for universe properties such as: what is the probability that the universe is large? 
Flat? 
Isotropic? 
Homogeneous?   
One obtains these via consideration of 
\beq
\mbox{Prob}(R) = \int_{R}|\Psi|^2\d\Omega 
\label{NSI}
\eeq
for $R$ a suitable region of the configuration space and $\d\Omega$ is the corresponding volume element.  
This approach is termed `na\"{\i}ve' due to it not using any further features of the constraint 
equations.  
The {\it conditional probabilities interpretation} \cite{PW83} goes further by addressing conditioned 
questions of `being' such as `what is the probability that the universe is flat given that it is 
isotropic'?  
{\it Records theory} \cite{PW83, GMH, B94II, EOT, H99, Records} involves localized subconfigurations 
of a single instant.
This requires notions of localization in space and in configuration space . 
One is furthermore in particular interested in whether these localized subconfigurations contain 
useful information and are correlated to each other (thus one needs notions of information, 
subsystem information, mutual information and so on), and whether this scheme leads to 
semblance of dynamics or history arising.   

\mbox{ }

\noindent D) Perhaps instead it is the histories that are primary ({\it histories theory} \cite{GMH, 
Histories}).    

\mbox{ }  

Further motivation along the lines of \cite{H03, FileR} for joining the semiclassical, 
histories and records approaches is as follows (see \cite{ASharp} for a more detailed account).    
This would both be a more robust Problem of Time strategy and useful for the investigation of a number 
of further issues in the foundations of Quantum Cosmology.  
The prospects of such a union are based on how, firstly, there is a records theory within histories theory. 
Secondly, decoherence of histories is one possible way of obtaining a semiclassical regime in the first 
place. 
Thirdly, what the records are will answer the further elusive question of which degrees of freedom  
decohere which others in Quantum Cosmology.

%========================================================================================================
\subsection{Quantum Cosmological motivation for the semiclassical approach} 
%========================================================================================================

The semiclassical approach is furthermore important toward acquiring more solid foundations for other 
aspects of Quantum Cosmology (see e.g. \cite{HT, H03, Kiefer99}). 
The abovementioned Halliwell-Hawking set-up amounts to our understanding of the quantum-cosmological 
origin of cosmological structure/small inhomogeneities.   
Via inflation in particular, a case can be built that Quantum Cosmology may contribute to our  
understanding of  cosmic microwave background fluctuations and the origin of galaxies \cite{Inf, HallHaw}.
Inflationary models have for the moment done well \cite{WMAP} at providing an explanation for these. 
This remains an `observationally active area', with the Planck experiment \cite{Planck} launched in 
2009.
Moreover, this paper aims at better understanding of the semiclassical approach itself, rather than 
looking to make any direct ties to observational cosmology.  
This is investigated qualitatively in this paper, using the following toy models.

%===========================================================================================================
\subsection{This paper uses relational particle models}
%===========================================================================================================

Scaled relational particle mechanics (RPM) (originally proposed in \cite{BB82} and further studied in 
\cite{B94I, EOT, 06I, TriCl, 08I, Cones, ScaleQM, 08III} is a mechanics in which only relative times, 
relative angles and relative separations have physical meaning.  
On the other hand, pure-shape RPM (originally proposed in \cite{B03} and further studied in \cite{06II, 
TriCl, FORD, 08I, 08II, +tri, AF, QShape, QSub, FileR}) is a mechanics in which only relative times, 
relative angles and ratios of relative separations have physical meaning.  
More precisely, these theories implement the following two Barbour-type (Machian) relational\footnote{RPM's 
%%%%%%%%%%%%%%%%%%%%%%%%%%%%%%%%%%%%%%%%%%%%%%%%%%%%%%%%%%%%%%%%%%%%%%%%%%%%%%%%%%%%%%%%%%%%%%%%%%%%%%%%%%%%%
are relational in Barbour's sense of the word rather than Rovelli's distinct one \cite{Rovellibook, B94I, 
EOT, 08I}.} 
%%%%%%%%%%%%%%%%%%%%%%%%%%%%%%%%%%%%%%%%%%%%%%%%%%%%%%%%%%%%%%%%%%%%%%%%%%%%%%%%%%%%%%%%%%%%%%%%%%%%%%%%%%%%%
postulates. 

\mbox{ }

\noindent 1) They are {\it temporally relational}.  
This means that there is no meaningful primary notion of time 
for the whole system thus described (e.g. the universe). 
This is implemented by using actions that are manifestly reparametrization invariant while also being 
free of extraneous time-related variables [such as Newtonian time or the lapse in General Relativity (GR)].   
This reparametrization invariance then directly leads to a primary constraint that is quadratic in the momenta. 
See Sec 2.2 for examples of such actions.  

\mbox{ } 

\noindent 2) They are {\it configurationally relational}.  
This can be thought of in terms of a certain group $G$ of transformations that act on the theory's 
configuration space $\fQ$ being held to be physically meaningless.   
One implementation of this uses arbitrary-$G$-frame-corrected quantities rather than `bare' 
$\fQ$-configurations.
For, despite this augmenting $\fQ$ to the principal bundle $P(\fQ, G)$, variation with respect to each 
adjoined independent auxiliary $G$-variable produces a secondary constraint linear in the momenta which 
removes one $G$ degree of freedom and one redundant degree of freedom from $\fQ$.   
Thus, one ends up on the desired reduced configuration space -- the quotient space $\fQ/G$.  
Configurational relationalism includes as subcases both spatial relationalism (for spatial 
transformations) and internal relationalism (in the sense of gauge theory).  
For scaled RPM, $G$ is the Euclidean group of translations and rotations, while for pure-shape RPM it 
is the similarity group of translations, rotations and dilations.  
Also see Sec 2 for actions for RPM's at various levels of reducedness, convenient 
coordinatizations and resulting quantum equations, 
which thus provides self-containedness and the notation for the paper.  
 
\mbox{ }

\noindent 
My principal motivation for studying RPM's\footnote{RPM's have elsewhere been motivated by the 
%%%%%%%%%%%%%%%%%%%%%%%%%%%%%%%%%%%%%%%%%%%%%%%%%%%%%%%%%%%%%%%%%%%%%%%%%%%%%%%%%%%%%%%%%%%%%%%%%%%%%%%%%
long-standing absolute or relational motion debate, and by RPM's making useful examples in the study of 
quantization techniques \cite{BS, RPMAPPS, Banal, FileR}. 
This paper's motivation follows from that in \cite{K92}. 
Moreover, I have now considerably expanded on this motivation by providing a very large number of 
analogies between RPM's and Problem of Time strategies. 
(See \cite{Cones, ASharp, FileR} for more detailed accounts of these analogies.)}
%%%%%%%%%%%%%%%%%%%%%%%%%%%%%%%%%%%%%%%%%%%%%%%%%%%%%%%%%%%%%%%%%%%%%%%%%%%%%%%%%%%%%%%%%%%%%%%%%%%%%%%%%
is that they are useful as toy models of GR in its traditional dynamical form (`geometrodynamics':  
the evolution of spatial geometries).  
The analogies between RPM's and GR (particularly in the formulations \cite{BSW+RWR} of 
geometrodynamics, c.f. Sec 2.2) are comparable in extent, but different to, the resemblance 
between GR and the more habitually studied minisuperspace models \cite{Mini}.  
{\sl RPM's are likely to be comparably useful as minisuperspace from the perspective of theoretical toy 
models}.
Some principal RPM--GR analogies are\footnote{See also Sec 2 for analogies at the level
%%%%%%%%%%%%%%%%%%%%%%%%%%%%%%%%%%%%%%%%%%%%%%%%%%%%%%%%%%%%%%%%%%%%%%%%%%%%%%%%%%%%%%%%%%%%%%%%%%%%%%%%%% 
of actions and of configuration spaces.}
%%%%%%%%%%%%%%%%%%%%%%%%%%%%%%%%%%%%%%%%%%%%%%%%%%%%%%%%%%%%%%%%%%%%%%%%%%%%%%%%%%%%%%%%%%%%%%%%%%%%%%%%%%

\mbox{ }

\noindent
1) the quadratic energy constraint\footnote{$\mbox{\bf R}^i$ are relative Jacobi 
%%%%%%%%%%%%%%%%%%%%%%%%%%%%%%%%%%%%%%%%%%%%%%%%%%%%%%%%%%%%%%%%%%%%%%%%%%%%%%%%%%%%%%%%%%%%%%%%%%%%%%%%%%%%
inter-particle (cluster) coordinates (see Sec 2 for more detail) with conjugate momenta $\mbox{\bf P}_i$. 
The $i$ and $j$ label $n$ = $N$ -- 1 relative separations of particles or particle clusters, where $N$ is 
the total number of particles.  
The corresponding configuration space metric is $M_{\alpha\beta ij} = \mu_i\delta_{ij}\delta_{\alpha\beta}$ 
where the $\mu_i$ are the corresponding cluster masses, with inverse denoted by $N^{\alpha\beta ij}$ and the 
lower-case Greek letters are spatial indices.   
$\fV$ denotes the potential energy and $\fE$ the total energy.}
%%%%%%%%%%%%%%%%%%%%%%%%%%%%%%%%%%%%%%%%%%%%%%%%%%%%%%%%%%%%%%%%%%%%%%%%%%%%%%%%%%%%%%%%%%%%%%%%%%%%%%%%%%%%
\beq 
\mH := N^{\alpha\beta ij}\mP_{\alpha i}\mP_{\beta j}/2 + \fV = \fE 
\eeq 
is the analogue of GR's quadratic Hamiltonian constraint (\ref{GRham}). 

\mbox{ }

\noindent
2) RPM's linear zero total angular momentum constraint 
\beq
\mbox{\bf L} := \sum\mbox{}_{\mbox{}_{\mbox{\scriptsize i}}} \mbox{\bf R}^i \cr \mbox{\bf P}_i = 0
\label{ZAM}
\eeq 
is a nontrivial analogue of GR's linear momentum constraint (\ref{GRmom}).

\mbox{ }

\noindent
3) In GR, 2) and the notion of local structure/clustering are tightly related as both concern the 
nontriviality of the spatial derivative operator.  
However, for RPM's, the nontriviality of angular momenta and the notion of 
structure/inhomogeneity/particle clumping are unrelated.  
Thus, even in the simpler case of 1-$d$ models, RPM's have nontrivial notions of structure 
formation/inhomogeneity/localization/correlations between localized quantities.  
In the subsequent 2-$d$ model paper \cite{08III}, one has nontrivial linear constraints too.
Each of these features is important for many detailed investigations in Quantum Gravity and Quantum 
Cosmology. 
Thus there are a number of specific ways in which RPM's, which possess nontrivial such features, 
are more useful than minisuperspace models, which do not.

\mbox{ }

RPM's are superior to minisuperspace for such a study as they have 

\mbox{ }

\noindent i) notions of localization in space. 

\mbox{ }

\noindent ii) They have more options for well-characterized localization in configuration space, i.e. of `distance 
between two shapes' \cite{NOD}.  
This is because RPM's have kinetic terms with positive-definite metrics, in contrast to GR's indefinite one.

\mbox{ }

\noindent iii) One can use RPM's to check the semiclassical approach's approximations and assumptions by using models 
that are exactly soluble by techniques outside the semiclassical approach.
One problem, which this paper builds further arguments for, is that a WKB regime cannot be expected to 
hold everywhere.  
Thus RPM's give a framework in which extra checks are possible as regards whether the WKB approximation 
holds well in all regions of interest. 

\mbox{ } 

\noindent 
iv) RPM's also have many further useful analogies \cite{K92, B94I, B94II, EOT, Paris, 06II, SemiclI, Records, 
08II, AF, Cones, ASharp} with GR at the level of conceptual aspects of Quantum Cosmology, including to 
many strategies for the Problem of Time.
As explained in the Conclusion, this is particularly the case for records theory, by which RPM's are 
a particularly suitable arena in which to investigate the unification of records, histories and semiclassical 
approaches.

%=====================================================================================================
\subsection{RPM model of the semiclassical approach and outline of the rest of this paper} 
%=====================================================================================================

\noindent Sec 3 sets up the semiclassical approach for this model.
This paper goes further than my previous semiclassical RPM paper \cite{SemiclI} via making use 
of a scale--shape split.  
It builds on the very brief \cite{MGM}, and includes the rather less trivial case of the relational 
triangle (for which \cite{08I, 08II, +tri, 08III} provides exact solution work).

\mbox{ }

I studied the semiclassical approach to RPM's before \cite{SemiclI}.  
The present paper's upgrades compared to that are  (as well as being models doable in the timeless and histories 
ways so as to permit comparison and composition of these approaches):

\mbox{ }

\noindent 
A) using scale coupled to a simpler but still nontrivial notion of shape than Halliwell and Hawking's.
N.B. that scaled RPM's in scale--shape split form with scale `heavy and slow' and shape `light 
and fast' make for more faithful models of semiclassical quantum cosmology than models with heavy, 
slow and light, fast particles.

\mbox{ } 

\noindent 
B) I do it for a reduced formulation (this is physically favoured since passing from reduced quantization 
to Dirac quantization is merely adding unphysical variables and so should not be capable of changing the physics, but it 
does lead to ambiguities in quantization procedure, and thus one should trust it less.)

\mbox{ } 

\noindent
C) I use the better-motivated conformal operator ordering (see Sec 2.4 and \cite{Banal}).    

\mbox{ } 

\noindent Sec 4 reviews \cite{Cones} cosmologically-inspired particular models and classical calculation of the 
heavy timefunction, the `rectifying timefunction' which  simplifies the emergent time dependent 
Schr\"{o}dinger equation and its inversion.

Also, in particular I set up 

\mbox{ }

\noindent
1)  A negligible-backreaction regime (Sec 5), involving a Hamilton--Jacobi equation and then an emergent-time-dependent 
perturbation of an emergent-time-dependent Schr\"{o}dinger equation. 

\mbox{ } 

\noindent 
2) A small-but-non-negligible backreaction regime (Sec 6), in which one encounters a Hamilton--Jacobi equation, then 
an emergent-time-dependent Schr\"{o}dinger equation, next an expectation-corrected Hamilton--Jacobi equation 
and finally a new emergent-time-dependent Schrodinger equation problem (with an inhomogeneous term based upon the 
lower-order wavefunction).  
2) is new to this paper. 

\mbox{ } 

\noindent 
In this paper, I solve both 1) and 2) modulo detail of the Hartree--Fock self-consistent scheme 
and leaving the last step in 2) in a formal form in terms of Green's functions.   
I comment on the former in the Conclusion (Sec 7), as well as delineating some interesting extensions of the present work 
to more complex models and to the tentative unification of semiclassical, histories and records approaches 
that Halliwell \cite{H99, H03, HT, H09} and I \cite{APOT, ASharp} are particularly interested in.

%========================================================================================================
%========================================================================================================
\section{Scaled RPM's}
%========================================================================================================
%========================================================================================================

%========================================================================================================
\subsection{Coordinatizations and reduced configuration spaces}
%========================================================================================================

Firstly I explain the what RPM's are in more detail, as well as the quantities used in their study.  
Absolutist configuration space $\fQ(N,d) = \mathbb{R}^{N d}$ is most conveniently coordinatized by 
$q^{\mu}_I$ $I$ = 1 to $N$, the particle number, and $\mu$ = 1 to $d$, the spatial dimension. 
Rendering absolute position irrelevant [e.g. by passing from particle position coordinates to any sort of 
relative coordinates i.e. taking out the centre of mass (COM) motion] leaves one on {\it relative 
configuration space} $\fR(N,d) = \mathbb{R}^{n d}$ for $n = N - 1$
In {\it relative Jacobi coordinates} \cite{Marchal} $R^{\mu}_i$  , the kinetic term is diagonal just as 
it was for the $q^{\mu}_I$, just for new values of the masses (see the next Subsection). 
The analogy with GR works well enough in dimension 2 (and more restrictedly so in dimension 1), so 
these are the cases that we consider.  
In dimension 2, one's model is an $N$-a-gonland (the smallest nontrivial such is triangleland), whilst 
in dimension 1 one's model is an `N-stop metroland'; this paper considers 3-stop metroland and 
triangleland examples.
E.g. for 3 particles, these are ${\bf R}_1$ and ${\bf R}_2$.  
These are combinations of relative position vectors ${\bf r}^{ab} = {\bf q}^b - {\bf q}^a$ 
between particles into inter-particle cluster vectors that are such that the kinetic term is cast 
in diagonal form: ${\bf R}_1 = {\bf q}_3 - {\bf q}_2$ and ${\bf R}_2 = {\bf q}_1 - \{m_2{\bf q}_2 + 
m_3{\bf q}_3\}/\{m_2 + m_3\}$.  
These have associated cluster masses $\mu_1 = m_2m_3/\{m_2 + m_3\}$ and $\mu_2 = 
m_1\{m_2 + m_3\}/\{m_1 + m_2 + m_3\}$.  
In fact, it is tidier as regards many of the paper's subsequent manipulations to use e.g. 
{\it mass-weighted} relative Jacobi coordinates ${\brho}^i = \sqrt{\mu_i}{\bf R}^i$ (Fig \ref{Fig1}). 
Physically, the squares of the magnitudes of these are the partial moments of inertia, $I^i = 
\mu_i|{{\bf R}^i}|^2$. 
%
%3) The normalized versions of 1), ${\bf n}^e := {\brho}^e/\rho$.   
%
Here, $\rho := \sqrt{I}$, the so-called {\it hyperradius}, and $I$ is the total moment of inertia. 
For specific components, I write the position indices downstairs as this substantially simplifies 
the notation.

I use (a) as shorthand for a,b,c forming a cycle along with the coordinatization being aligned with 
the clustering (i.e. partition into subclusters) in which bc form a pair and a is a loose particle 
(e.g. the split of the three vertices of a triangle into a base pair and an apex).
I take clockwise and anticlockwise labelled triangles to be distinct, and particles to be distinguishable.  
I.e. I make the plain rather than mirror-image-identified choice of set of shapes with labelled vertices; 
I do so for simplicity - I strongly desire simple maths in order to take many Problem of Time calculations far enough to 
consider adjoining/unifying them and simple maths that is quantum-cosmologically-interpretable is essential for this 
and that is precisely what small RPM models provide.

%FFFFFFFFFFFFFFFFFFFFFFFFFFFFFFFFFFFFFFFFFFFFFFFFFFFFFFFFFFFFFFFFFFFFFFFFFFFFFFFFFFFFFFFFFFFFFFFFFFFFFFF
{            \begin{figure}[ht]
\centering
\includegraphics[width=0.6\textwidth]{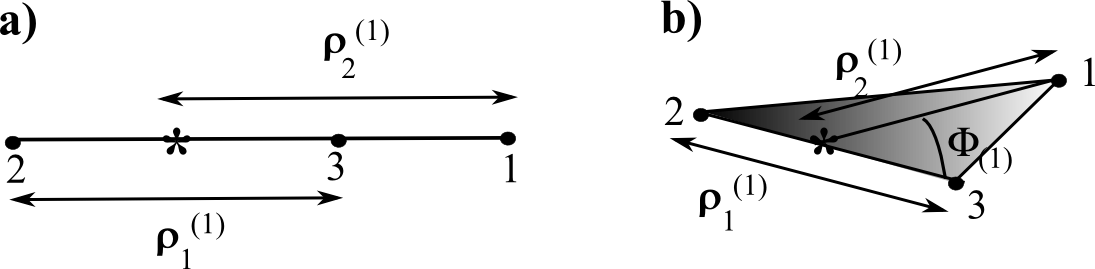}
\caption[Text der im Bilderverzeichnis auftaucht]{        \footnotesize{
\noindent
a) For 3 particles in 1-$d$, one permutation of relative Jacobi coordinates is as indicated. 
* denotes the centre of mass of particles 2 and 3.  

\noindent 
b) The same permutation of relative Jacobi coordinates for 3 particles in 2-$d$. 
In {\sl mass-weighted space}, the magnitudes are the length of a base of the triangle and what is a 
median in the equal-mass case. 
I define the `Swiss army knife' angle between the two $\brho_1^{(\sa)}$ by $\Phi_{(\sa)} = 
\mbox{arccos}\big({\brho}_1^{(\sa)}\cdot{\brho_2^{(\sa)}}/\rho_1^{(\sa)}\rho_2^{(\sa)}\big)$,  
and the ratio variable ${\Theta}_{(\sa)} = \mbox{arctan}(\rho_1^{(\sa)}/\rho_2^{(\sa)})$.    }         }
\label{Fig1}\end{figure}          }
%FFFFFFFFFFFFFFFFFFFFFFFFFFFFFFFFFFFFFFFFFFFFFFFFFFFFFFFFFFFFFFFFFFFFFFFFFFFFFFFFFFFFFFFFFFFFFFFFFFFFFFF
 
\mbox{ }

\noindent 
If rotation with respect to absolute axes is to have no meaning, then one is left on a configuration 
space {\it relational space} ${\cal R} = \mathbb{R}^4/SO(2)$. 
If instead absolute scale were to have no meaning, then one is left on a configuration space 
\cite{Kendall} {\it preshape space} = $\mathbb{R}^4/\mbox{Dil}$ (for Dil the dilational group). 
It is straightforward to see that this is $\mathbb{S}^{3}$. 
If both of the above are to have no meaning, then one is left on \cite{Kendall} {\it shape space},  
$\fS = \mathbb{R}^4/SO(2) \times \mbox{Dil}$.   
Taking Relative space to correspond to the space of Riemannian 3-metrics on a fixed spatial topology $\Sigma$  
(taken to be compact without boundary for simplicity), then relational space corresponds to Wheeler's superspace($\Sigma$) 
\cite{Wheeler}, shape space to conformal superspace CS($\Sigma$) and preshape space to pointwise conformal superspace \cite{FM96}.
Finally the relational configuration space is the cone over the shape space.  
At the topological level, for C(X) to be a cone over some topological manifold X, 
\beq
\mbox{C(X) = X $\times$ [0, $\infty$)/\mbox{ }$\widetilde{\mbox{ }}$} \mbox{ } , 
\eeq
where the meaning of $\mbox{ } \widetilde{\mbox{ }}$ is that all points of the form \{p $\in$ X, 0 $\in 
[0, \infty)$ \} are `squashed' i.e. identified to a single point termed the {\it cone point}, and 
denoted by 0. 
At the level of Riemannian geometry, a cone C(X) over a Riemannian space X possesses 
a) the above topological structure and 
b) a Riemannian line element given by 
\beq
\d S^2 = \d{\cal R}^2 + {\cal R}^2\d s^2 \mbox{ } .
\eeq
Here, $\d s^2$ is the line element of X itself and ${\cal R}$ is a suitable `radial variable' that 
parametrizes the [0, $\infty$), which is the distance from the cone point.  
This metric is smooth everywhere except (possibly) at the troublesome cone point.

Now, $\mC(\mathbb{S}^2)$ is, at the topological level, $\mathbb{R}^3$.  
However, this $\mathbb{R}^3$ and $\mathbb{S}^2$ are not straightforward realizations at the level of 
configuration space metric geometry. 
The $\mathbb{R}^3$ has a curved metric on it and a dimensionally unintuitive radial variable, 
as follows.
The shape space sphere turns out to have radius 1/2, as can be seen from the relational space line 
element\footnote{$p$, $q$, $r$ are relational space indices (in this paper, these run from 1 to 3).  
%%%%%%%%%%%%%%%%%%%%%%%%%%%%%%%%%%%%%%%%%%%%%%%%%%%%%%%%%%%%%%%%%%%%%%%%%%%%%%%%%%%%%%%%%%%%%%%%%%%%%%%%%
$u$, $v$, $w$ are shape space indices (in this paper's triangleland case, these run form 1 to 2, whilst 
in 3-stop metroland there is just one shape space coordinate).  
I also use straight indices (upper or lower case) to denote quantum numbers, the index $\mS$ to denote 
`shape part' and the index $\rho$ (referring to the hyperradius) to denote `scale part'.}
%%%%%%%%%%%%%%%%%%%%%%%%%%%%%%%%%%%%%%%%%%%%%%%%%%%%%%%%%%%%%%%%%%%%%%%%%%%%%%%%%%%%%%%%%%%%%%%%%%%%%%%%%  
\beq
\d S^2 = \d\rho^2 + \rho^2\{\d\Theta^2 + \mbox{sin}^2\Theta\,\d\Phi^2\}/4 \mbox{ } , \mbox{ } \mbox{ } \mbox{corresponding to kinetic 
metric} 
\mbox{ }  
{\cal M}_{pq} = \mbox{diag}(1, \rho^2/4, \rho^2\mbox{sin}^2\Theta/4) \mbox{ } . 
\eeq
This inconvenience in coordinate ranges is then overcome by using the moment of inertia $I$ instead as the radial variable, 
\beq
\d S^2 = \{\d I^2 + I^2\{\d\Theta^2 + \mbox{sin}^2\Theta\,\d\Phi^2\}\}/4I 
\mbox{ } , \mbox{corresponding to } 
{\cal M}_{pq} = \mbox{diag}({1}/{4I}, \mbox{ } {I}/{4}, \mbox{ } {I\mbox{sin}^2\Theta}/{4}) 
\mbox{ } .
\eeq 
This metric ${\cal M}_{pq}$ is not the usual flat metric on $\mathbb{R}^3$: it is curved.  
However, it is clearly conformal to the flat metric
\beq
\d S^2_{\sF\sll\sa\st} = \d I^2 + I^2\{\d\Theta^2 + \mbox{sin}^2\Theta\,\d\Phi^2\} 
\mbox{ } , \mbox{corresponding to } 
\overline{{\cal M}}_{pq} = \mbox{diag}
\left(
1, \mbox{ } I^2, \mbox{ } I^2\mbox{sin}^2\Theta
\right) \mbox{ } . 
\eeq
(This is in spherical polar coordinates with $I$ as radial variable, the conformal factor relating it to the 
previous metric being $\Omega^2 = 1/4 I$, a fact that is subsequently exploited in this paper). 
That $I$ features as radial variable is the start of significant differences between the triangleland 
and 4-stop metroland configuration spaces (the latter having the more intuitively obvious $\rho$ as 
radial variable).

The 1-$d$ and 2-$d$ cases of this have shape spaces $\mathbb{S}^{N - 2}$ and $\mathbb{CP}^{N - 2}$ 
respectively, including the configuration space line elements (the coned ones, containing the 
unconed ones as $\d s^2_{\sss\sp\sh\se}$ and $\d s^2_{\sF\sS}$ that are subsequently spelt out 
(FS stands for the Fubini--Study metric on $\mathbb{CP}^{N - 2}$, see e.g. 
\cite{FORD, Cones, QShape, QSub} for how this arises in the $N$-a-gonland mechanics context).

%========================================================================================================
\subsection{Actions for RPM's}
%========================================================================================================

These follow from Jacobi-type actions \cite{Lanczos} 
\beq
\fS = 2\int\d\lambda\sqrt{\fT\{\fU + \fE\}} \mbox{ } ,
\label{action}
\eeq
where the kinetic term $\fT$ is, in the particle position presentation, 
\beq
\fT = \sum\mbox{}_{\mbox{}_{\mbox{\scriptsize $I$}}}
\frac{    m_I\{\dot{\mq}^{I\alpha}  - \dot{\ma} - \{\dot{\mb}\cr\dot{\mq}^I\}^{\alpha}\}
             \{\dot{\mq}^I_{\alpha} - \dot{\ma} - \{\dot{\mb}\cr\dot{\mq}^I\}_{\alpha}\}     }{2} = 
\frac{    m_I\delta_{IJ}\delta_{\alpha\beta}
             \{\dot{\mq}^{I\alpha} - \dot{\ma} - \{\dot{\mb}\cr\dot{\mq}^I\}^{\alpha}\}
             \{\dot{\mq}^{J\beta}  - \dot{\ma} - \{\dot{\mb}\cr\dot{\mq}^J\}^{\beta} \}      }{2} 
                                                        \mbox{ } 
\eeq
(using the Einstein summation convention in the second expression, and where $\ma^{\alpha}$ 
and $\mb^{\alpha}$ are translational and rotational auxiliary variables.  
In the relative Jacobi coordinates presentation, it is 
\beq
\fT = \sum\mbox{}_{\mbox{}_{\mbox{\scriptsize $i$}}} 
\frac{    \mu_i\{\dot{\mR}^{i\alpha} - \{\dot{\mb}\cr\dot{\mR}^i\}^{\alpha}\}
               \{\dot{\mR}^i_{\alpha} - \{\dot{\mb}\cr\dot{\mR}^i\}_{\alpha}\}    }{2} = 
\frac{    \mu_i\delta_{ij}\delta_{\alpha\beta}
               \{\dot{\mR}^{i\alpha} - \{\dot{\mb}\cr\dot{\mR}^i\}^{\alpha}\}
               \{\dot{\mR}^{j\beta} - \{\dot{\mb} \cr\dot{\mR}^j\}^{\beta}\}      }{2} 
                                                        \mbox{ } .
\eeq
In the reduced formulation for 3-stop metroland, 
\beq
\fT = \{\dot{\rho}^2 + \rho^2\dot{\varphi}^2\}/2 
\eeq
where $\varphi = \mbox{arctan}(\rho_2/\rho_1)$.  
In the reduced formulation for triangleland, 
\beq
\fT = \frac{1}{2}
\left\{
\dot{\rho}^2 + \frac{\rho^2}{4}\{\dot{\Theta}^2 + \mbox{sin}^2\Theta\dot{\Phi}^2\}
\right\} = \frac{1}{2}\frac{1}{4I}
\left\{
\dot{I}^2 + I^2\{\dot{\Theta}^2 + \mbox{sin}^2\Theta\dot{\Phi}^2\}
\right\} \mbox{ } . 
\eeq
These actions implement temporal relationalism via reparametrization invariance: $\fT$ is purely quadratic in $\d/\d\lambda$ and 
occurs as a square root factor so that this $\d/\d\lambda$ cancels with the $\d\lambda$ of the integration and thus the 
$\lambda$ is indeed a mere label.
These actions implement configurational relationalism via the corrections to the $\dot{\mq}_I$ and $\dot{\mR}_i$ 
(the linear constraint coming from variation with respect to $\mb^{\alpha}$ is \ref{ZAM}) and via 
using G-invariant constructs directly in the reduced approach.     
The GR counterpart, for comparison and to further substantiate the tightness of the GR--RPM analogy is an also reparametrization-invariant 
and thus temporal relationalism implementing action that is a variant of the Baierlein--Sharp--Wheeler action \cite{BSW+RWR},   
\beq
\fS = \int\d\lambda\int\sqrt{h}\sqrt{\fT_{\sG\sR}\{- 2\Lambda + R\}}
\eeq
where
\beq
\fT_{\sG\sR} = M^{\mu\nu\rho\sigma}\{\dot{h}_{\mu\nu}     - \pounds_{\dot{\sF}}h_{\mu\nu}    \}
                                   \{\dot{h}_{\rho\sigma} - \pounds_{\dot{\sF}}h_{\rho\sigma}\} \mbox{ } .  
\eeq
Here, $M^{\mu\nu\rho\sigma}$ is the undensitized version of the GR configuration space metric (of which the 
DeWitt supermetric is the inverse), equal to $h^{\mu\rho}h^{\nu\sigma} - h^{\mu\nu}h^{\rho\sigma}$,
$\dot{\mF}$ plays the same role mathematically as the GR shift and 
$\pounds_{\dot{\sF}}$ is the the Lie derivative with respect to $\dot{\mF}$.
The $\pounds_{\dot{\sF}}h_{\mu\nu}$ corrections to the metric velocities then indirectly implement 
configurational relationalism with respect to the 3-diffeomorphisms; the associated constraint is (\ref{GRmom}).

%======================================================================================
\subsection{Momenta, constraints and conserved quantities}
%======================================================================================

We are interested in the reduced case in the triangleland case in which this is different from the unreduced/Dirac approach.   
For 3-stop metroland, the momenta are
\beq
p_{\rho} = *{\rho} , \mbox{ } \mbox{ } p_{\varphi} = \rho^2*{\varphi} \mbox{ } .
\eeq
(Here, $* := \sqrt{\fT/\{\fE - \fV\}}\,\dot{\mbox{}}$. 
There is also a single energy constraint
\beq
\{p_{\rho}^2 + p_{\varphi}^2/\rho^2\}/2 + \fV = \fE \mbox{ } .
\eeq
For triangleland, the momenta are
\beq
p_{I} = *{I} , \mbox{ } \mbox{ } p_{\theta} = I^2*{\theta} \mbox{ } ,  \mbox{ }  \mbox{ }  
p_{\phi} = I^2\mbox{sin}^2\theta*\phi \mbox{ } .  
\eeq
[now $\overline{*} = \sqrt{\fT/\{\fE - \fV\}}\,\dot{\mbox{}}$ for `banal-conformally' redefined $\overline{\fT} = \Omega^2 T$ and 
$\overline{\fE} - \overline{\fV} = \Omega^{-2}\{\fE - \fV\}$ in which the $\Omega^2 = 1/4I$ conformal factor has been passed 
over from the $\fT$ factor of the action to the $\fE - \fV$ factor, which clearly leaves the action (\ref{action}) invariant)].  
There is now also a single energy constraint
\beq
\{p_{I}^2 + \{p_{\theta}^2+ p_{\phi}^2/\mbox{sin}^2\theta\}/I^2\}/2 + \fV = \fE \mbox{ } .
\eeq
For the classical equations for this, see \cite{08I, ScaleQM} and for the kinematical quantization, see \cite{AF, ScaleQM, +tri, 08III}.

\mbox{ }

In 3-stop metroland, $\varphi$-independent potentials have a conserved quantity ${\cal D}$, 
which is to be interpreted as the relative dilational momentum of the 2 constituent subsystems 
(the particle pair and the third particle).   
Mathematically, dilational momentum is the dot product counterpart of angular momentum's cross product; 
physically it is indeed associated with change of size i.e. dilation alias expansion.    
In triangleland, $\Phi$-independent potentials have a conserved quantity ${\cal J}$ that is the relative 
angular momentum of the two constituent subsystems.
$\Phi$ and $\Theta$ independent potentials have in addition to ${\cal J}$ two conserved quantities ${\cal R}_1$ and ${\cal R}_2$, 
which have the mathematics of angular momenta [associated with the isometry group SO(3) of the triangeland 
configuration space sphere); however, physically, the ${\cal R}_1$ and ${\cal R}_2$ are mixtures of relative angular momenta 
and relative dilational momenta (and had been previously called {\sl generalized} angular momenta \cite{Smith}].  
From this mathematical correspondence, clearly the sum of the squares of these, ${\cal T}\!\mo\mt = {\cal R}_1^2 + {\cal R}_2^2 + {\cal J}^2$ 
will also be significant. 
I also refer to ${\cal D}^2$ as ${\cal T}\!\mo\mt$ in the 3-stop metroland context.

%========================================================================================================
\subsection{Corresponding timeless Schr\"{o}dinger equations}
%========================================================================================================

The timeless Schr\"{o}dinger equation for the general scaled RPM is as follows. 
I present it for the $\xi$-operator ordering, making also use of the formula $D^2\Psi = \rho^{1 - \sn d} \pa_{\rho} 
\{\rho^{\sn d - 1} \pa_{\rho}\Psi\} + \rho^{-2}D_{\sS}^2\Psi$ for $D_{\sS}^2$ the Laplacian 
on shape space, which follows from the geometrical considerations in \cite{Cones}. 
[In fact, I favour the conformal-ordered case among these \cite{Banal}, but this makes little difference as regards 
this paper's semiclassical workings, so I keep $\xi$ general.  
The motivation for the family of $\xi$-orderings is that they remain invariant under changes of coordinatization of 
the configuration space.  
Among these, the Laplacian ordering ($\xi = 0$) is the simplest, whilst the conformal ordering preserves a conformal 
symmetry that turns out to be the same as the banal-conformal symmetry of the relational actions of Sec 2.2.]
Then for 1-$d$ RPM or 2-$d$ RPM in $\mathbb{CP}^k$ presentation henceforth collectively referred to as case A),
\beq
-\hbar^2\{\rho^{1 - \sn d}\pa_{\rho}\{\rho^{\sn d - 1}\pa_{\rho}\Psi\} + \rho^{-2}
\{D^2_{\sS}\Psi - \kappa(\xi)\Psi\}\} + 2\fV(\rho)\Psi + 2\fJ(\rho, \mS^u)\Psi = 2\fE\Psi \mbox{ }   
\eeq
[c.f. the Wheeler-DeWitt equation (\ref{WDE}) for GR].
I also split the potential $\fV$ into heavy part $\fV(\rho$ alone) and light-and-interaction part 
$\fJ(\rho, \mS^{\sfA})$. 
(This splitting is taken to include the approximate case, in which writing shape terms as an expansion 
can give a nontrivial shape independent lead term for but small fluctuations in shape).
[Above, $\kappa(\xi)$ a constant equal to $\xi \, \mbox{Ric}(M)$, which is 0 for C($\mathbb{S}^{\sn - 1}$) = 
$\mathbb{R}^n$ with flat metric and $6 \, n \, \xi$ for C($\mathbb{CP}^{\sn - 1}$). 
In the conformal ordered case, this becomes $3n\{2n - 3\}/4\{n - 1\}$. ]

\mbox{ }

Additionally, for triangleland in the $\mathbb{S}^2$ presentation [henceforth referred to as case B)]
\beq
-\hbar^2\{I^{2}\pa_{\rho}\{I^{2}\pa_{\rho}\Psi\} + I^{-2}
\{{D}^2_{\sS}\Psi - \kappa(\xi)\Psi\}\} + 2\overline{\fV}(I)\Psi + 2\overline{\fJ}(I, \mS^u)\Psi = E/2I\Psi \mbox{ } .  
\eeq
using the banal-conformally-transformed $\overline{\fV} = \overline{\fV}(I) + \overline{\fJ}(I, \mS^u)$.  

\vspace{3in}

%========================================================================================================
\section{Semiclassical approach to the shape-scale split reduced RPM}
%========================================================================================================

I present this for case A); for case B) instead, use barred quantities and $I$ in place of $\rho$.
The Introduction's general case here involves scale $\rho$ being heavy and slow and shape $\mS^{u}$ being 
light and fast.   
The Born--Oppenheimer ansatz is then 
\beq
\Psi(\rho, \mS^{u}) = \psi(\rho)|\chi(\rho, \mS^{u})\rangle \mbox{ } ,  
\label{BO}
\eeq
and the WKB ansatz is
\beq
\psi(\rho) = \mbox{exp}(i \fW(\rho)/\hbar) \mbox{ } . 
\label{WKB}
\eeq
There is then the issue of approximations associated with each of these as detailed in \cite{SemiclI} 
for the simplest operator ordering and in \cite{FileR} for the case in hand.   
Various pieces of `folklore' as regards some of these approximations are exposed in Secs 4.5 and 5.5   
While one is accustomed to seeing WKB procedures in ordinary QM, N.B. that these rest on the 
`Copenhagen' presupposition that one's quantum system under study has a surrounding classical large 
system and that it evolves with respect to an external time.   
Moreover, in Quantum Cosmology, as the quantum system is already the whole universe, the notions of a 
surrounding classical large system and of external time cease to be appropriate \cite{Wheeler, Zeh}.  
So using WKB procedures in Quantum Cosmology really does require novel and convincing justification, 
particularly if one is relying on it to endow a hitherto timeless theoretical framework with 
a bona fide emergent time.
Were this attainable, it would then go a long way toward rigorously resolving the `paradox' in the 
sense that the truly relevant procedure of inspection of $l$-subsystems would reveal a semblance of 
dynamics even if the universe is, overall, timeless. 

\noindent Next, the $h$-equation $\langle\chi| \times$ (time-independent Schr\"{o}dinger equation), with the 
associated integration being over the $l$ degrees of freedom and thus over shape space, is
\beq
\langle\chi| {\cal O} |\chi\rangle = \int_{\sfS(N, d)}\chi^{\star} \, {\cal O} \,\chi\, \fD \mS
\eeq 
for $\fD \mS$ the measure over shape space) with the ans\"{a}tze (\ref{BO}) and (\ref{WKB}) substituted in gives
$$ 
\{\pa_{\rho}\fW\}^2 - i\hbar\pa_{\rho}\mbox{}^2\fW - 2i\hbar\pa_{\rho}\fW\langle\chi|\pa_{\rho}|\chi\rangle - 
\hbar^2\{  \langle\chi|\pa_{\rho}\mbox{}^2|\chi\rangle + 
           \{n d - 1\}\rho^{-1}\langle\chi|\pa_{\rho}|\chi\rangle   \} - 
i\hbar\rho^{-1}\{n d - 1\}\pa_{\rho}\fW + \hbar^2\rho^{-2}k(\xi) 
$$
\beq
+ 2\fV_{\rho}(\rho) + 
2\langle\chi|\fJ(\rho, \mS^{u})|\chi\rangle = 2\fE \mbox{ } \mbox{ } .  
\eeq
Here, I have discarded an additional $\hbar^2\langle\chi|D^2|\chi\rangle$ term by integration by 
parts and the shape spaces in question being compact without boundary, and the WKB approximation has 
additionally killed off the second term. 

\noindent
Also, (time-independent Schr\"{o}dinger equation) -- $|\chi\rangle \times$ (h equation) gives the $l$-equation
\beq
\{1 - P_{\chi}\}\{- 2i\hbar\pa_{\rho}|\chi\rangle\pa_{\rho}W + 
\hbar^2\{\pa_{\rho}\mbox{}^2|\chi\rangle + \pa_{\rho}|\chi\rangle \{n d - 1\}\rho^{-1} + \rho^{-2}D^2_{\sS}\} + 
J(\rho, \mS^{u})|\chi\rangle\} = 0  
\eeq
for $P_{\chi}$ the projector $|\chi\rangle\langle\chi|$.

Now let us use $\pa^2_{\rho} \fW$ negligible (WKB approximation) and apply 
\beq
\pa_{\rho}\fW = p_{\rho} = *\rho
\label{lance}
\eeq
by the expression for momentum in the Hamilton--Jacobi formulation and the momentum-velocity relation,  
and using the chain-rule to recast $\pa_{\rho}$ as $\pa_{\rho}t\,*$.
Then 
$$ 
\{*\rho\}^2 - i\hbar\pa_{\rho}t**\rho - 2i\hbar *\rho\langle\chi|\pa_{\rho}t*|\chi\rangle - 
\hbar^2\{  \langle\chi|\{\pa_{\rho}t*\}^2|\chi\rangle + 
           \{n d - 1\}\rho^{-1}\langle\chi|\pa_{\rho}t*|\chi\rangle   \} - 
i\hbar\rho^{-1}\{n d - 1\}*\rho + \hbar^2\rho^{-2}k(\xi) 
$$
\beq
+ 2\fV_{\rho}(\rho) + 
2\langle\chi|J(\rho, \mS^{u})|\chi\rangle = 2\fE \mbox{ } \mbox{ } .  
\label{TimeSet}
\eeq

The first form of $h$-equation collapses to (`neglecting $\hbar^2$ terms and averages') 
the Hamilton--Jacobi equation   
\beq
\{\pa_{\rho}\fW\}^2 = 2\{\fE - \fV_{\rho}\} \mbox{ } , 
\label{Sicut}
\eeq
whilst the second form of $h$-equation likewise collapses to the corresponding energy equation 
\beq
\rho^{*\,2} = 2\{\fE - \fV_{\rho}\} \mbox{ } .
\label{Dixit} 
\eeq
A reformulation of this of use in further discussions in this paper is the analogue Friedmann 
equation, 
\beq
\left\{\frac{\rho^*}{\rho}\right\}^2 = \frac{2\fE}{\rho^2} - \frac{2\fV_{\rho}}{\rho^{2}} \mbox{ } .
\label{Vaire}
\eeq
This equation is also later explicitly required for case B): 
\beq
\left\{\frac{\overline{*}{I}}{I}\right\}^2 =  \frac{\fE}{2I^3} - \frac{{\cal T}}{I^4} 
                                                     - \frac{\fV(I, \, \mS^u)}{2I^3} \mbox{ } .
\label{Este}
\eeq                                              
The energy equation form is then solved by 
\beq
t^{\se\sm} - t^{\se\sm}_0 = \frac{1}{\sqrt{2}}\int\frac{\d\rho}{\sqrt{\fE - \fV_{\rho}}} \mbox{ } \mbox{ or } \mbox{ } \mbox{ }
t^{\se\sm} - t^{\se\sm}_0 = {\sqrt{2}}\int\frac{\sqrt{I}\d I}{\sqrt{\fE - \fV_{I}}} \mbox{ } .  
\label{Jesu}
\eeq

\noindent
Next, there is a cross-term move to obtain a time-dependent Schr\"{o}dinger equation: the first term in 
($l$-equation)$/2$ is $i\hbar\pa|\chi\rangle/\pa t^{\se\sm}$ by (\ref{lance}) and the chain-rule in 
reverse:
\beq
N^{\rho\rho}i\hbar\frac{\pa W}{\pa \rho}\frac{\pa|\chi\rangle}{\pa \rho} = 
i\hbar N^{\rho\rho}\pi\frac{\pa|\chi\rangle}{\pa \rho} =
i\hbar N^{\rho\rho}M_{\rho\rho}*\rho\frac{\pa|\chi\rangle}{\pa \rho} = 
i\hbar \frac{\pa \rho}{\pa t}\frac{\pa|\chi\rangle}{\pa \rho}  = 
i\hbar \frac{\pa|\chi\rangle}{\pa t}\mbox{ } .  
\eeq
Thus, one obtains the $l$-equation in the fluctuation form, 
\beq
\{1 - P_{\chi}\}i\hbar \frac{\pa |\chi\rangle}{\pa t} = \{1 - P_{\chi}\}
\left\{
-\frac{\hbar}{2} 
\left\{
\frac{1}{\rho(t)^2}D^2_{\sS} |\chi \rangle + 
\frac{\pa t}{\pa \rho} \frac{\pa }{\pa t} 
\left\{
\frac{\pa t}{\pa \rho}\frac{\pa }{\pa t}
\right\}|\chi\rangle + 
\frac{n d - 1}{\rho}\frac{\pa t}{\pa \rho} \frac{\pa }{\pa t} 
|\chi\rangle + \fJ|\chi\rangle
\right\}
\right\} \mbox{ } .  \label{Fluc}
\eeq

Then use (\ref{Sicut}), (\ref{Dixit}) or (\ref{Jesu}) to express $\rho$ as a function of $t^{\se\sm}$.  
N.B. it is logical and consistent for the other $\rho$-derivatives to also be expressed as $t$-derivatives, giving in full: 
$$
\{1 - P_{\chi}\}i\hbar\frac{  \pa|\chi\rangle  }{  \pa t^{\se\sm}  } = \{1 - P_{\chi}\}
\left\{
-\frac{\hbar^2}{2}
\left\{
\frac{    1    }{    \rho^2(t^{\se\sm})    }{D}^2_{\sS} |\chi\rangle + 
\frac{    1    }{    \sqrt{  2\{\fE - \fV_{\rho}(\rho(t^{\se\sm}))\}  }    }\frac{\pa}{\pa t^{\se\sm}}
\left\{
\frac{    1    }{    \sqrt{  2\{\fE - \fV_{\rho}(\rho(t^{\se\sm}))\}  }    }
\frac{\pa|\chi\rangle}{\pa t^{\se\sm}}
\right\}
\right.
\right.
$$
\beq
\left.
\left.
+ \frac{n d - 1}{\rho(t^{\se\sm})}
\frac{    1    }{    \sqrt{2\{\fE - \fV_{\rho}(\rho(t^{\se\sm}))\}  }    }
\frac{\pa|\chi\rangle}{\pa t^{\se\sm}}
\right\}
+ \fJ|\chi\rangle
\right\} \mbox{ } .  
\eeq
Now consider (\ref{TimeSet}) and (\ref{Fluc}) as a pair of equations to solve for the unknowns $t^{\se\sm}$ and $|\chi\rangle$. 
One often neglects these extra $t^{\se\sm}$-derivative terms whether by discarding them prior to noticing they are also convertible 
into $t^{\se\sm}$-derivatives or by arguing that $\hbar^2$ is small or $\rho$ variation is slow. 
Moreover there is a potential danger in ignoring higher derivative terms even if they are small (c.f. Navier--Stokes equation  
versus  Euler equation in the fluid dynamics)
\noindent
We need the invertibility in order to set up the $t^{\se\sm}$-dependent perturbation equation and more generally 
have a time provider equation followed by an explicit time-dependent rather than heavy degree of freedom dependent equation.

%====================================================
\subsection{Rectified time}
%====================================================

The time-dependent Schr\"{o}dinger equation core 
\beq
i\hbar\frac{\pa|\chi\rangle}{\pa t^{\se\sm}} = - \frac{\hbar^2}{2\rho^2(t^{\se\sm})}D^2_{\sS} |\chi\rangle + \fJ|\chi\rangle 
\eeq
simplifies if one furthermore chooses the rectified time given by [for case A)] 
\beq
\rho^2\pa/\pa t^{\se\sm} = \pa/\pa t^{\sr\se\sc} \mbox{ } , 
\eeq
i.e. 
\beq
t^{\sr\se\sc} - t^{\sr\se\sc}(0) =  \int\d t^{\se\sm}/\rho^2(t^{\se\sm}) \mbox{ } .  
\eeq
We can get to this all in one go by combining its definition with the energy equation (so as 
to cancel out the emergent time): 
\beq
t^{\sr\se\sc} - t^{\sr\se\sc}(0) = \frac{1}{\sqrt{2}}\int \frac{\d \rho}{\rho^2\sqrt{{\fE - \fV(\rho)}}} \mbox{ } . 
\label{Tu}
\eeq
On the other hand, for case B), we use instead the `rectified time' \cite{Cones}
\beq
t^{\sr\se\sc} - t^{\sr\se\sc}(0) = \int\d t^{\se\sm}/I(t^{\se\sm})^2 \mbox{ } .  
\eeq 
Then, all in one go, 
\beq
t^{\sr\se\sc} - t^{\sr\se\sc}(0) = \sqrt{2}\int  
\frac{\d I}{I^{2}}\sqrt{\frac{I}{\fE - \fV(I)}} \mbox{ } .  
\label{Rek}
\eeq

As regards interpreting the rectified timefunction, in each case using $t^{\sr\se\sc}$ amounts to 
working on the shape space itself, i.e. using the geometrically natural presentations of Sec 3.1.   
Note: to keep formulae later on in the paper tidy, I use $t$ for $t^{\se\sm}$ and ${\cal T}$ for $t^{\sr\se\sc}$ 
up to a constant [always including $t^{\sr\se\sc}(0)$ and sometimes including the constant of integration from 
the other side of (\ref{Tu})].  
Finally, approximately isotropic GR has an analogue of rectification too, amounting to absorption of 
extra factors of the scalefactor $a$ viewed as a function of $t^{\se\sm}$.  

%\mbox{ } 

\noindent {\bf Lemma}: suppose $t^{\se\sm}$ is monotonic.  
Then the rectified time ${\cal T}$ is also monotonic.

\mbox{  }

\noindent\underline{Proof} 
For case A), 
\beq
\frac{\d {\cal T}}{\d \rho} = \frac{\d {\cal T}}{\d t^{\se\sm}}\frac{\d t^{\se\sm}}{\d \rho} = 
\frac{1}{\rho(t^{\se\sm})^2}\frac{\d t^{\se\sm} }{\d\rho} \geq 0
\eeq
by the chain-rule in step 1, (\ref{Tu}) in step 2 and the positivity of squares and the 
assumed monotonicity of $t^{\se\sm}$ in step 3.
For case B), the $\rho \longrightarrow I$, barred counterpart of this argumentation holds also $\Box$.

\mbox{ }

In terms of this, one can view the $l$-equation as (perhaps perturbations about) a time-dependent 
Schr\"{o}dinger equation on the shape space, 
\beq
i\hbar\frac{\pa|\chi\rangle}{\pa {\cal T}} = \frac{-\hbar^2}{2}D^2_{\sS}|\chi\rangle + 
\widetilde{\fJ}|\chi\rangle \mbox{ \{+ further perturbation terms\} }
\eeq
(this paper's specific examples of which are, mathematically, familiar equations). 
$\widetilde{\fJ}$ denotes $\rho^2(t^{\se\sm}({\cal T}))\fJ$.

\mbox{ } 

\noindent The rectified time's simplification of the emergent-time-dependent Schr\"{o}dinger equation 
can be envisages as passing from the emergent time that is natural to the whole relational space to 
a time that is natural on the shape space of the $l$-degrees of freedom themselves, i.e. to working 
on the shape space of the $l$-physics itself.

%=================================================================================================================
\subsection{Perturbation series solution of the system}
%=================================================================================================================

Parametrize the smallness of the interaction term between the $h$ and $l$ subsystems by splitting out a factor $\epsilon$.
I.e. $\fJ \longrightarrow \epsilon \fJ$.  
Apply then $t = t_{(0)} + \epsilon t_{(1)}$ and $|\chi\rangle = |\chi_{(0)}\rangle + \epsilon|\chi_{(1)}\rangle$.

%=================================================================================================================
\subsection{The negligible-backreaction regime}
%=================================================================================================================

Usually we keep the first $\fJ$, since elsewise the $l$-system's energy changes without the $h$-system responding, 
violating conservation of energy.    
But if this is just looked at for a ``short time" (few transitions, the drift may not be great, and lie within the 
uncertainty to which an internal observer would be expected to know their universe's energy.
Then the system becomes
\beq
\left\{
\frac{\pa \rho}{\pa t_{(0)}}
\right\}^2 
= 2\{\fE - \fV\}  \mbox{ } ,   
\label{Manwe}
\eeq
\beq
i\hbar\frac{\pa|\chi\rangle}{\pa {\cal T}_{(0)}} = -\frac{\hbar^2}{2}D^2_{\sS}|\chi\rangle + \epsilon \widetilde{\fJ}|\chi\rangle  \mbox{ } .
\label{Varda}
\eeq
Here, we do not explicitly perturbatively expand the last equation as it is a decoupled problem of a standard form:
a ${\cal T}$-dependent perturbation of a simple and well-known ${\cal T}$-dependent perturbation equation.
I provide solutions in some specific cases of these equations in Sec 5.

%====================================================================================================
\subsection{Extension}
%====================================================================================================

Including expectation terms in equation (\ref{Varda}) then gives a Hartree--Fock self-consistent scheme.
This is $t$-dependent for sure, and not involving an antisymmetrized wavefunction, but nevertheless it still is a 
known set-up 

\mbox{ } 

\noindent{\bf Open question 1}: variationally justify this Hartree--Fock self-consistent procedure, 
and then study the outcome of it.

%=======================================================================================
\subsection{Small but non-negligible backreaction regime}
%=======================================================================================

We now look to solve 
\beq
\frac{1}{2}
\left\{
\frac{\pa\rho}{\pa t}
\right\}^2 + 
\epsilon\langle\chi|\,\fJ\,|\chi\rangle = \fE - \fV \mbox{ } , 
\eeq
\beq
i\hbar\frac{\pa|\chi\rangle}{\pa t} = - \frac{\hbar^2}{2\rho^2(t)}D_{\sS}^2|\chi\rangle + \epsilon \fJ|\chi\rangle \mbox{ } . 
\eeq
[Note that if there were a separate $\fV_S$, rectification leads to this becoming part of $\widetilde{\fJ}$, and, in any case, 
we are considering $\fV$ 's that are homogeneous in scale variable, thus I do not write down a separate $\fV_{\sS}$: the isotropic  
$\fV_{\rho}$ and the direction-dependent interaction term $\fJ$ contain everything that ends up to be of relevance.]

In the small but non-negligible-backreaction regime, there is a $t_{(1)}$ equation to solve, and the $|\chi_{(1)}\rangle$ stage then gives 
the following set of equations 
\beq
\d\rho^2 = 2\{\fE - \fV(\rho)\}\d t_{(0)}^2 \mbox{ } ,
\eeq
\beq
i\hbar\frac{\pa|\chi_{(0)}\rangle}{\pa {\cal T}_{(0)}} = -\frac{\hbar^2}{2}D^2_{\sS}|\chi_{(0)}\rangle \mbox{ } ,
\eeq
\beq
\d t_{(1)}\{\fE - \fV(\rho)\} = \langle\chi_{(0)}|\,\fJ\,|\chi_{(0)}\rangle \d t_0 \mbox{ } ,
\label{three}
\eeq
\beq
i\hbar\frac{\pa|\chi_{(1)}\rangle}{\pa {\cal T}_{(0)}} = -\frac{\hbar^2}{2}D^2_{\sS}|\chi_{(1)}\rangle + 
\left\{ 
\widetilde{\fJ} - \frac{\hbar^2}{2}\frac{\d {\cal T}_{(1)}}{\d {\cal T}_{(0)}}D^2_{\sS}
\right\}
|\chi_{(0)}\rangle
\mbox{ } .
\eeq 

\noindent
One can then solve (\ref{three}) for $t_{(1)}$ using knowledge of the solutions of 
the first two decoupled equations, 
\beq
t_{(1)} - t_{(1)}(0) = \int_{t_{0}(0)}^{t_{0}}
\left\{
\frac{1}{2\big\{\fE - \fV\big(\rho_{(0)}\big(t_{(0)}^{\prime}\big)\big)\big\}}\int \chi_{(0)}^*(t, \mS^u)\,\fJ\,\chi(t, \mS^u) \fD S \d t_{(0)}^{\prime}
\right\}
\eeq
and then also using 
\beq
\frac{\d {\cal T}_{(1)}}{\d {\cal T}_{(0)}} = \frac{\pa t_{(1)}}{\pa t_{(0)}} = 
\frac{1 }
                    {2\big\{\fE - \fV\big(\rho_0\big(t_{(0)}^{\prime}\big)\big)\big\}}\int \chi_{(0)}^*\,\fJ\,\chi_{(0)} \fD \mS
\eeq
to render the fourth equation into the form 
\beq
i\hbar\frac{\pa|\chi\rangle}{\pa {\cal T}_{(0)}} = -\frac{\hbar^2}{2}D^2_{\sS}|\chi_{(1)}\rangle 
+ \left\{
\widetilde{\fJ} - \frac{\hbar^2\langle\chi_{(0)}|\,\fJ\,|\chi_{(0)}\rangle D_{\sS}^2}{4\big\{\fE - \fV\big(\rho\big(t_{(0)}\big)\big)\big\}}
\right\}|
\chi_{(0)}\rangle \mbox{ } . 
\label{Flaot}
\eeq
I provide partial solutions to this negligible-backreaction scheme in some specific cases in Sec 6. 
[Modulo leaving the solution of (\ref{Flaot}) in a formal form involving the below use of Green's functions.]  

\noindent 
Note: the last term with the big bracket is by this stage a known, so this is just an inhomogeneous version of the 
second equation and therefore amenable to the method of Green's functions.
$$
|\chi_{(1)}\rangle = \int_{{\cal T}^{\prime} = 0}^{{\cal T}}\int_{\sfS(N, d)}
G(\mS^u, {\cal T}_{(0)}; \mS^{u\,\prime}, {\cal T}_{(0)}^{\prime})
\left\{
\widetilde{\fJ} - \frac{\hbar^2\langle \chi_{(0)}|\,\fJ\,|\chi_{(0)}\rangle D^2_{\sS}  \rangle}{4\big\{\fE - \fV\big(\rho\big(t^{\prime}_{(0)}
\big({\cal T}_{(0)}^{\prime}\big)\big)\big)  \big\}} 
\right\}
|\chi(\mS^{u\,\prime}, {\cal T}_{(0)}^{\prime})\rangle \, \fD \mS^{\prime}\d {\cal T}^{\prime} 
$$
\beq
=: \int_{{\cal T}^{\prime} = 0}^{{\cal T}}\int_{\sfS(N, d)}
G(\mS^u, {\cal T}_{(0)}; \mS^{u\,\prime}, {\cal T}_{(0)}^{\prime})f(\mS^{u\prime}, {\cal T}^{\prime}) \fD\mS^{\prime} \d{\cal T}^{\prime}
\label{Munch}
\eeq
modulo additional boundary terms/complementary function terms.
Now, this is a very standard linear operator for this paper's simple RPM examples (time-dependent 1-$d$ and 2-$d$ rotors).
However, i) the region in question is less standard (an annulus or spherical shell with the time variable playing the role of  
radial thickness).  
ii) Nor is it clear what prescription to apply at the boundaries. 
On these grounds, I do not for now provide explicit expressions for these Green's functions, though this 
should be straightforward enough once ii) is accounted for.

\mbox{ } 

\noindent {\bf Open question 2}:  it is not as yet clear how to extend the self-consistent treatment to this non-negligible backreaction case.
It is something like an emergent-time-dependent Hartree--Fock scheme coupled to a time-producing expectation-corrected variant of the 
Hamilton--Jacobi equation.

%=================================================================================================================
%=================================================================================================================
\section{Specific examples of small RPM models at the classical level}
%=================================================================================================================
%=================================================================================================================

%========================================================================================================
\subsection{RPM -- Cosmology analogy}
%========================================================================================================

The models are to be interpreted according to the following analogy.  
The energy equations cast in the form (\ref{Vaire}) for case A) or (\ref{Este}) 
for case B) are analogous to the Friedmann equation 
\beq
\left\{\frac{1}{a}\frac{\d a}{\d t^{\sc\so\sss}}\right\}^2 = 
-\frac{k}{a^2} + \frac{\Lambda}{3}  + \frac{8\pi G\varepsilon}{3} \mbox{ } . 
\eeq
(for $t^{\sc\so\sss}$ the cosmic time) {\sl after} use 
of the energy-momentum conservation equation or the Raychaudhuri equation to turn the energy density $\varepsilon$
The scalefactor $a$ corresponds to  the hyperradius scale $\rho$ for the $N$-stop metroland and the 
moment of inertia scale $I$ for triangleland.

\mbox{ }  

\noindent In this paper, I consider (sums of) power-law potentials; these are motivated by being are common in mechanics 
and by their mapping to the commonly-studied terms in the Friedmann equation of cosmology also.  
E.g. for 3-stop metroland,  
\beq
\fV = C_1|\mq_2 - \mq_3|^{\zeta} + C_2|\mq_3 - \mq_1|^{\zeta} + C_3|\mq_1 - \mq_2|^{\zeta}
\label{char}
\eeq
becomes 
\beq
\fV = C_1\rho^{\zeta}|\mbox{cos}\,\varphi|^{\zeta} + 
      C_2\rho^{\zeta}|\mbox{cos}\,\varphi + \sqrt{3}\mbox{sin}\,\varphi|^{\zeta} + 
      C_3\rho^{\zeta}|\mbox{cos}\,\varphi - \sqrt{3}\mbox{sin}\,\varphi|^{\zeta}  \mbox{ } .  
\label{3stoppot}
\eeq
On the other hand, for triangleland, (\ref{char}) becomes
\beq
\fV = C_1\rho^{\zeta}|\mbox{sin}\mbox{$\frac{\Theta}{2}$}|^{\zeta} + 
C_2\rho^{\zeta}|\mbox{cos}\mbox{$\frac{\Theta}{2}$} - \mbox{$\frac{1}{2}$}\mbox{sin}\mbox{$\frac{\Theta}{2}$}|^{\zeta} + 
C_3\rho^{\zeta}|\mbox{cos}\mbox{$\frac{\Theta}{2}$} + \mbox{$\frac{1}{2}$}\mbox{sin}\mbox{$\frac{\Theta}{2}$}|^{\zeta} 
\mbox{ } .  
\eeq
HO-type potential RPM models are then exactly soluble, which motivates them as highly tractable.
More widely, these analogies with cosmology require that any shape factors present being slowly-varying so that one 
can carry out the following scale-dominates-shape approximation holding at least in some region of interest. 
\beq
|\fT_{l}| << |\fT_{h}| \mbox{ realized by the shape quantity } \fT_{\sS} <<   \fT_{\rho} 
\mbox{  (scale quantity) } ,
\label{SSA1}
\eeq
\beq
|\fJ_{hl}| << |\fV_{h}| \mbox{ realized by } |\fJ(\rho, \mS^{u})| << |\fV_{(0)}(\rho)|  
\label{SSA2}
\eeq
for $\fJ_{hl}$ the interaction part of the potential.  
\beq
\fV(\rho, \mS^{u}) \approx \fV_{(0)}(\rho)\{1 + \fV_{(1)u}(\rho)\mS^{u} + 
O(|\mS^{u}|^2)\} = \fV_0(\rho) + \fJ(\rho, \mS^{u}) \mbox{ } .  
\label{SSA3}
\eeq 
[Without this, one cannot separate out the heavy (here scale) part so that it can provide the approximate 
timefunction with respect to which the light (here shape) part's  dynamics runs.
Moreover, N.B. that isotropic cosmology itself similarly suppresses small anisotropies and inhomogeneities, 
so that exact solutions thereof are really approximate solutions for more realistic universes too.]
Or the barred, $\rho \longrightarrow I$ counterpart of this for the $\mathbb{S}^2$ presentation of triangleland.

\mbox{ }  

The rest of the Cosmology--RPM analogy is as follows.
Cosmology's spatial curvature term $k$ is paralleled by $-2E$ in case A) and $2A$ in case B).  
To match the $k$ = --1, 0 or 1 convention, I use the corresponding scaling freedom in RPM's to 
set $-2E$ =  --1, 0 or 1 in the standard analogy, and $2A$ = --1, 0 or 1 in the $\mathbb{S}^2$ 
presentation of triangleland's analogy (which freedom remained unused in \cite{08I, 08II, +tri}); this 
amounts to the other coefficients being redefined by that constant factor; of course, this does not 
change any results). 
The  cosmological constant term $\Lambda/3$ is paralleled by $-2A$ in case A) and the surviving term from 
$r_{IJ}^6$ potentials.
%
%, with coefficient $2\sigma$ in case B).
%
Cosmology's dust term coefficient $2GM$ is paralleled by $-2K$ from the Newtonian potentials in case A) 
and by $2E$ in case B). 
Cosmology's radiation term coefficient $2GM$ is paralleled by $2R$ from conformally-invariant potential in 
both case A) and case B). 
In each case also ${\cal T}\!\mo\mt$ contributes as radiation but with the wrong sign, so that the overall term is 
$2R - {\cal T}\!\mo\mt$.
\noindent
While the wrong-sign radiation term is not a conventional term in Cosmology, 
one can get it in e.g. brane cosmology, however, due to bulk effects); 
moreover, in the present context this term has an obvious analogy with ordinary mechanics: the centripetal term 
which acts to prevent collapse to the origin (which, in the cosmological context, is a bounce solution).

%========================================================================================================
\subsection{Some GR isotropic cosmology solutions}
%========================================================================================================

Most of this paper's analogue-cosmology models correspond to the following standard cosmological models.
The first four are for 3-stop metroland.

\mbox{ }

\noindent i)   $k = -1$ and $\Lambda > 0$ is $a = \sqrt{3/\Lambda}\,\mbox{sinh}(\sqrt{\Lambda/3}t^{\sc\so\sss})$: 
de Sitter/inflationary type model with negative curvature, corresponding to an RPM with upside-down HO 
potentials and positive energy.  

\mbox{ }

\noindent
ii)  $k = 1$ and $\Lambda > 0$ is $a = \sqrt{3/\Lambda}\,\mbox{cosh}(\sqrt{\Lambda/3}t^{\sc\so\sss})$:   
de Sitter/inflationary type model with positive curvature, corresponding to an RPM with upside-down HO 
potentials and negative energy.  

\mbox{ }

\noindent
iii) $k = -1$ and $\Lambda < 0$ is $a = \sqrt{-3/\Lambda}\,\mbox{sin}(\sqrt{-\Lambda/3}t^{\sc\so\sss})$ 
-- a `Milne in anti de Sitter' \cite{Rindler} oscillating solution, corresponding to an RPM with HO 
potentials and positive energy.  

\mbox{ }

\noindent
For these quadratic potentials, $\fV  =\{K_1\rho_1^2 + K_2\rho_2^2\}/2$ for Hooke's coefficients 
$K_1$ and $K_2$ becomes $\fV = \rho^2\{K_1\mbox{sin}^2\varphi + K_2\mbox{cos}^2\varphi\}/2 = 
\rho^2\{\{K_1 + K_2\}/4 + \{K_2 - K_1\}\mbox{cos}2\varphi/4 := \rho^2\{A + B\mbox{cos}2\varphi\}$.
Thus A is an isotropic HO coefficient and B is a measure of contents inhomogeneity for the model universe.  

\mbox{ }

\noindent iv) $k = 0$, $\Lambda = 0$ is $a = \{9GM/2\}^{1/3}t^{^{\sc\so\sss}\,2/3}$ for the case of dust matter, corresponding to
an RPM with Newtonian potentials and zero energy.  

\mbox{ }  

\noindent
v) is a rather simpler but analytically realizable the more complicated and in some ways more useful 
triangleland context: the $k = -1$, $\Lambda = 0$ cosmology, for which $a = t^{\sc\so\sss}$, corresponding to, in the 
RPM, upside-down HO potentials and zero energy.  

\mbox{ } 

N.B. a wide range of other more realistic cosmological solutions, e.g. involving also dust and radiation matter 
are realizable as RPM analogue models as outlined in \cite{ScaleQM, 08III}.  
Models i)--iii) and v) are selected as simple examples that remain particularly analytically tractable in doing 
emergent semiclassical time strategy to the Problem of Time calculations.  
iv) is used in further discusions.  
These calculations are however (more laboriously and making use of more approximations) also doable in these 
more realistic models.  
I currently consider this paper's range of models to be sufficient, however, for the principal aims of 
qualitatively assessing Halliwell and Hawking and looking into unifications of semiclassical, records and 
histories strategies.

%===========================================================================================================
\subsection{$t^{\se\sm}$ for each RPM counterpart of these}
%===========================================================================================================

Now take each of these solutions under the analogy, and then invert the scale variable--$t^{\se\sm}$  relation.  
(In all the cases considered, this invertibility exists and is analytically tractable). 
For model i), $t^{\se\sm} = \{1/\sqrt{-2A}\} \mbox{arsinh}(\sqrt{-2A}\rho)$ + const. 
For model ii), $t^{\se\sm} = \{1/\sqrt{-2A}\} \mbox{arcosh}(\sqrt{-2A}\rho)$ + const.
For model iii), $t^{\se\sm} = \{1/\sqrt{2A}\}  \mbox{arcsin}(\sqrt{2A}\rho)$ + const.  
For model iv), $t^{\se\sm} = \sqrt{-2K/9}\rho^{3/2}$ + const.
For model v), one has $t^{\se\sm} = I$ + const. 
I note that all of the above approximate heavy-scale timefunctions are monotonic apart from 
model iii), which nevertheless has a reasonably long era of monotonicity as regards modelling early-universe 
Quantum Cosmology.  
Model iii) has periods proportional to $1/\sqrt{A}$ for $N$-stop metroland, which plays a role proportional 
to that of $1/\sqrt{\Lambda}$ in GR Cosmology.  
%
%Also, ${\cal A}$)iv) hits zero scale at other than t = 0 (one might reset origin of time to deal with 
%this).
%
Finally, for model ii) there is a nonzero minimum size.

%=================================================================================================
\subsection{Rectified time for each of the preceding}
%=================================================================================================

Now let us make use of the rectified time variable ${\cal T}$ so as to be able to pass to the 
simplified Schr\"{o}dinger equation on the shape space.   
In all of the cases considered (both here and 
more extensively in \cite{ScaleQM, 08III}) this is analytically possible, and invertible so that 
one can make the scale a function of the rectified time analytically, by composition of two 
analytical inversions.  
For model i)   ${\cal T} = - \mbox{coth}(\sqrt{-2A}t^{\se\sm})/\{-2A\}^{3/2} = -\sqrt{1 - 2A\rho^2}/4A^2\rho$, 
for model ii)  ${\cal T} =   \mbox{tanh}(\sqrt{-2A}t^{\se\sm})/\{-2A\}^{3/2} = \sqrt{-2A\rho^2 - 1}/4A^2\rho$ and 
for model iii) ${\cal T} = - \mbox{cot}(  \sqrt{2A}t^{\se\sm}  )/\{  \sqrt{2A}\}^{3/2} = -\sqrt{1 - 2A\rho^2}/4A^2\rho $. 
For model iv), ${\cal T} = - \{-2/K\}^{2/3}/3t^{\se\sm\, 1/3} = \{-2/K\}^{2/3}/3\{-2K/9\}^{1/6}\rho^{1/2}$.  
For model v),
\beq
{\cal T} = \sqrt{\frac{2}{-A}}\left\{\frac{1}{I_0} - \frac{1}{I}\right\} \mbox{ } . 
\eeq
Thus, inverting, $I = 1/\{1/I_0 - \sqrt{-A/2}{\cal T}\} = \sqrt{2/-A}/\{\check{\cal T} - {\cal T}\}$ for 
$\check{\cal T} := \sqrt{2/-A}/I_0$.

%===============================================================================================================
\subsection{Regions in which various approximations apply}
%===============================================================================================================

For 3-stop metroland, (\ref{3stoppot}) is stable to small angular disturbances about $\varphi$ for some 
cases of harmonic oscillator/cosmological constant, but it is unstable to small angular disturbances 
for the gravity/dust sign of inverse-power potential. 
Positive-power potentials are finite-minimum wells, but cease to be exactly soluble.    
For Newtonian gravity/dust models (or negative power potentials more generally), near the corresponding 
lines of double collision, the potential has abysses or infinite peaks. 
Thus the scale-dominates-shape approximation, which represents a standard part of the approximations 
made in setting up the semiclassical approach to Quantum Cosmology, fails in the region around these lines.  
It is also then possible that dynamics that is set up to originally run in such a region leaves it, so a 
more detailed stability analysis is required to ascertain whether semiclassicality is stable in such models.   
This issue can be interpreted as a conflict between the procedure used in the semiclassical approach and the 
example of trying to approximate a 3-body problem by a 2-body one (see \cite{ScaleQM} for further comments on 
this).  
The difference in the analogy between Cosmology and case B) causes the 
potentials for that which are studied in this paper (and the wider range of these considered in \cite{08III}) 
to be purely positive powers for which the preceding problem does not occur.

%===================================================================================================
\section{Negligible backreaction QM scheme}
%===================================================================================================

%===================================================================================================
\subsection{First approximation for time-dependent wave equation for 3-stop metroland}
%===================================================================================================

(\ref{Varda}) can be viewed as a ${\cal T}$-dependent perturbation of what is, in the $N$-stop metroland 
case, a well-known ${\cal T}$-dependent Schr\"{o}dinger equation (the usual one on the 
circle/sphere/hypersphere).

\mbox{ }

For the 3-stop metroland HO case, 
this time-dependent Schr\"{o}dinger equation is [for cases i) to iii)]
\beq
i\hbar\frac{\pa|\chi\rangle}{\pa {\cal T}} = - \frac{\hbar^2}{2M}\frac{\pa^2|\chi\rangle}{\pa\varphi^2} + 
\frac{  B\,\mbox{cos}\,2\varphi  }{4A^2\{1 + \{2|A|\}^3{\cal T}^2\}^2  }  |\chi\rangle 
\mbox{ } ,
\label{main}
\eeq
which, for $B$ small in relation to e.g. the $A$-term or the $D^2_{\fS}$ term and corresponding to $\epsilon$ small in this example, 
i.e. $K_1 \approx K_2$, poses, about a very simple quantum equation, a (fairly analytically tractable) ${\cal T}$-dependent perturbation problem.  
The reason for studying the above `negative curvature balanced by negative cosmological constant' 
type scenario is that it is exactly soluble by means not usually available in semiclassical approach studies, 
allowing for a number of further checks (see \cite{FileR}).  

\mbox{ }

Note that many of the other examples listed in \cite{Cones} can be taken as far as an analytic counterpart of eq. (\ref{main}) 
time-dependent Schr\"{o}dinger equation. 
However, I omit these further examples due to complications in subsequent stages of the working.

%========================================================================================================
\subsection{Perturbation theory integrals for 3-stop metroland}
%========================================================================================================

The unperturbed equation is solved by
\beq
|\chi^{\sc(0)}_{\sd}\rangle = \mbox{exp}(i\fE_{\sd}{\cal T}/\hbar)\mbox{cos}(\md\varphi)/2\pi \mbox{ } , \mbox{ } \mbox{ }   
|\chi^{\sss(0)}_{\sd}\rangle = \mbox{exp}(i\fE_{\sd}{\cal T}/\hbar)\mbox{sin}(\md\varphi)/2\pi
\eeq
for $\fE_{\sd} = \hbar^2\md^2/2$.  
 
\mbox{ }

The standard approach to time-dependent perturbation theory gives, to first order [(1)-superscripts]    
\beq
\langle \psi^{\st(1)}_{\sd^{\prime}}| \psi_{\sd}^{\st(0)} \rangle = \delta_{\sd\sd^{\prime}} - 
\frac{iB}{4\hbar A^2} \, I^{\st}(\md,\md^{\prime}) \, I(\beta, {\cal T}) 
\eeq
for the following split-up integrals (the superscript t stands for `trig function' and takes the 
values c for the cosine solution and s for the sine solution).  
\beq
I^{\sc}(\md,\md^{\prime}) := \int \frac{\mbox{cos}\,\md^{\prime}\varphi}{\sqrt{2\pi}} \mbox{cos}\,2\varphi 
                \frac{\mbox{cos}\,\md \varphi}{\sqrt{2\pi}} \d\varphi = 
                \frac{1}{4}\left\{\delta_{\sd^{\prime} + 2,\sd} + \delta_{\sd^{\prime} - 2,\sd}\right\} 
= \int \frac{\mbox{sin}\,\md^{\prime}\varphi}{\sqrt{2\pi}} \mbox{cos}\,2\varphi 
  \frac{\mbox{sin}\,\md \varphi}{\sqrt{2\pi}} \d\varphi =: I^{\sss}(\md,\md^{\prime})
\eeq
for d, $\md^{\prime} >$ 2, the other cases including the following nonzero exceptions:   
\beq
I^{\sc}(0, 2) = I^{\sc}(2, 0) = 1/2 \mbox{ } , \mbox{ }     \mbox{ } 
I^{\sc}(1, 1) = 1/4                           \mbox{ }   \mbox{ and } \mbox{ } \mbox{ } 
I^{\sss}(1, 1) = -1/4                         \mbox{ } .   
\eeq
Also, 
\beq
I(\beta, {\cal T}) := \int_{0}^{{\cal T}}\mbox{exp}(i\beta{\cal T}^{\prime} ) 
\d{\cal T}^{\prime}/\{1 + 8|A |^3{\cal T}^{\prime\,2}\}^2 \mbox{ },   
\eeq
where 
\beq
\beta := \hbar\{\md^{\prime\,2} - \md^2 \}/2 \mbox{ } .  
\eeq
This integral is simple in the $\beta = 0$ (d = $\md^{\prime}$) case (the only survivor of which by the $\phi$-integral is 
d = $\md^{\prime}$ = 1)  for which the complex numerator collapses to 1: 
\beq
I(0, {\cal T}) = \frac{1}{4\sqrt{2}A^{3/2}}
\left\{
\frac{2\sqrt{2}A^{3/2}{\cal T}}{8A^3{\cal T}^2 + 1} + \mbox{arctan}(2\sqrt{2}A^{3/2}{\cal T}) \mbox{ } .
\right\}
\eeq
For all the other cases, the integral is complex, and comes out as a complicated combination of elementary, Si and 
Ci functions that I do not provide.  

\mbox{ } 

The first-order perturbed wavefunctions then comes out as 
\beq
|\psi^{\st(1)}_{d}({\cal T}) \rangle = | \psi^{\st(0)}_{\sd} \rangle - \frac{iB}{16\hbar A^2}
\{ I(2\hbar\{1 + d\}, {\cal T}) | \psi^{\st(0)}_{\sd + 2} \rangle   +  
\{ I(2\hbar\{1 - d\}, {\cal T}) | \psi^{\st(0)}_{\sd - 2} \rangle  \} \mbox{ } \md > 2 \mbox{ } ,
\eeq
\beq
|\psi^{\sc(1)}_{2}({\cal T}) \rangle = | \psi^{\sc(0)}_{2} \rangle   - \frac{iB}{16\hbar A^2}
\{I(6\hbar, {\cal T}) | \psi^{\sc(0)}_{4} \rangle   + 
2 I(-2\hbar, {\cal T}) | \psi^ {\sc(0)}_0 \rangle  \} \mbox{ }  , 
\eeq
\beq
|\psi^{\sc(1)}_{1}({\cal T}) \rangle = | \psi^{\sc(0)}_{1} \rangle - \frac{iB}{16\hbar A^2}
I(0, {\cal T}) | \psi^{\sc(0)}_{1} \rangle   \mbox{ } ,
\eeq
\beq
|\psi^{\ss(1)}_{1}({\cal T}) \rangle = | \psi^{\ss(0)}_{1} \rangle + \frac{iB}{16\hbar A^2}
I(0, {\cal T}) | \psi^{\ss(0)}_{1} \rangle   \mbox{ } , \mbox{ and }
\eeq
\beq
| \psi^{\sc(1)}_{0}({\cal T}) \rangle = | \psi^{\ss(0)}_{1} \rangle - \frac{iB}{8\hbar A^2}
I(4\hbar, {\cal T}) | \psi^{\ss(0)}_{2} \rangle  \mbox{ }  .
\eeq

%===========================================================================================
\subsection{Time-dependent wave equations for triangleland}
%===========================================================================================

\beq
i\hbar\pa_{\cal T}|\chi\rangle = -\{\hbar^2/2\}D_{\sS}^2|\chi\rangle + 
\widetilde{\fJ}({\cal T}, \sS^{u})|\chi\rangle \mbox{ } \mbox{ for } \mbox{ }  
\widetilde{\fJ} = I^2(t^{\se\sm}({\cal T}))^2 \fJ \mbox{ } ,  
\eeq
is also a ${\cal T}$-dependent perturbation of a well-known ${\cal T}$-dependent 
Schr\"{o}dinger equation (the usual one on the sphere).
For model v)'s potential, one has the time-dependent Schr\"{o}dinger equation
\beq
i\hbar\frac{\pa|\chi\rangle}{\pa {\cal T}} = -\frac{\hbar^2}{2}
\left\{
\frac{1}{\mbox{sin}\Theta}\frac{\pa}{\pa \Theta}\{\mbox{sin}\Theta\frac{\pa}{\pa\Theta}  + \frac{\pa^2}{\pa\varphi^2}
\right\}
|\chi\rangle + 
\frac{B\,\mbox{cos}\,\Theta}{4A\{{\cal T} - \check{\cal T}\}^2}|\chi\rangle  \mbox{ } . 
\eeq

%========================================================================================================
\subsection{Perturbation theory integrals for this example}
%========================================================================================================

The unperturbed equation is solved by
\beq
|\chi^{\sc(0)}_{\sd}\rangle = \mbox{exp}(i\fE_{\sR}{\cal T}/\hbar)Y_{\sR\sj}(\Theta, \Phi)
\eeq
for $\fE_{\sR} = \hbar^2\mR^2/2$ and $Y_{\sR\sj}$ the usual spherical harmonics (albeit with unusual 
meanings for the quantum numbers paralleling Sec 2.3's conserved quantity discussion and unusual 
interpretations of the spherical angles as per Sec 2.1).

The standard approach to time-dependent perturbation theory gives, to first order [(1)-superscripts];    
\beq
\langle\psi_{\sR^{\prime}\sj^{\prime}}| \psi_{\sR\sj}({\cal T})\rangle = 
\delta_{\sR^{\prime}\sR}\delta_{\sj^{\prime}\sj} + \frac{iB}{2\hbar A} 
I(\gamma, {\cal T}) I_{\Theta\Phi}(\mR,\mR^{\prime},\mj,\mj^{\prime})
\eeq
for 
\beq
\gamma := \hbar\{\mR^{\prime \, 2} - \mR^2\}/2  
\eeq
and for the following split-up integrals.
Firstly, 
\beq
I_{\Theta\Phi}(\mR,\mR^{\prime},\mj,\mj^{\prime}) := 
\langle\Psi_{\sR^{\prime}\sj^{\prime}} | \mbox{cos}\,\Theta | \Psi_{\sR\sj}\rangle 
\eeq
which has the same mathematical form as the integral that appears in the derivation of the selection 
rules for electric dipole transitions \cite{AF, BlumMessiah}; more generally, it is a simple 
example of a 3-$Y$ integral \cite{Mizu}.  
Its nonzero cases are given by 
\beq
\langle \psi_{\sR + 1, \sj} |\mbox{cos}\,\Theta|\psi_{\sR \sj} \rangle =  
\sqrt{\frac{\{\mR + 1\}^2 - \mj^2}{\{2\mR + 1\}\{2\mR + 3\}}}
\label{first} \mbox{ } , 
\eeq
and 
\beq
\langle \psi_{\sR - 1, \sj} |\mbox{cos}\,\Theta |\psi_{\sR \sj} \rangle = \sqrt{\frac{\mR^2 - \mj^2}{\{2\mR - 1\}\{2\mR + 1\}}}  \mbox{ } . 
\eeq
Secondly, 
$$
I(\gamma, {\cal T}) := \int_0^{{\cal T}}
{\d{\cal T}^{\prime}}\mbox{exp}(i\gamma{\cal T}^{\prime})/\{{\cal T}^{\prime} - \check{\cal T}\}^2 =
$$
\beq
\frac{\mbox{exp}(i\beta{\cal T})}{\check{\cal T} - {\cal T}} - \frac{1}{\check{\cal T}} + 
\gamma
\left\{ 
 \mbox{sin}\,\gamma\check{\cal T}\{\mbox{Ei}(i\gamma\{\check{\cal T} - {\cal T}\}) - \mbox{Ei}(i\gamma\check{\cal T}) - 
i\mbox{cos}\,\gamma\check{\cal T}\{\mbox{Ei}(i\gamma\{{\cal T} - \check{\cal T}\}) - \mbox{Ei}(-i\gamma\check{\cal T})
\right\} \mbox{ } .
\eeq
The final answer for the first-order perturbed wavefunctions then takes the form 
$$
|\psi_{\sR\,\sj}({\cal T})\rangle \propto |\psi_{\sR\,\sj}\rangle + 
\frac{i}{2\hbar}\frac{B}{A}
\left\{
\left\{
\frac{\mbox{exp}(i\hbar\{\mR + 1\}{\cal T})}{\check{\cal T} - {\cal T}} - \frac{1}{\check{\cal T}} + 
\hbar\{\mR + 1\}
\left\{ 
\mbox{sin}(\hbar\{\mR + 1\}\check{\cal T})
\{
\mbox{Ei}(i\hbar\{\mR + 1\}\{\check{\cal T} - {\cal T}\}) 
\right.
\right.
\right.
$$
$$
- 
\mbox{Ei}(i\hbar\{\mR + 1\}\check{\cal T})
\} - 
i\mbox{cos}(\hbar\{\mR + 1\}\check{\cal T})
\{
\mbox{Ei}(i\hbar\{\mR + 1\}\{{\cal T} - \check{\cal T}\}) - 
\mbox{Ei}(-i\hbar\{\mR + 1\}\check{\cal T})
\} 
\left\}
\left\}
\sqrt{\frac{\{\mR + 1\}^2 - \mj^2}{\{2\mR + 1\}\{2\mR + 3\}}}|\psi_{\sR + 1,\sj}\rangle +
\right.
\right.
$$
$$
\left. 
\left\{
\frac{\mbox{exp}(-i\hbar\mR{\cal T})}{\check{\cal T} - {\cal T}} - \frac{1}{\check{\cal T}} -  
\hbar\mR 
\left\{ 
-\mbox{sin}(\hbar\mR\check{\cal T})
\{
\mbox{Ei}(i\hbar\mR\{{\cal T} - \check{\cal T}\}) - 
\mbox{Ei}(-i\hbar\mR\check{\cal T})
\} - 
\right.
\right.
\right.
$$
\beq
\left.
i\mbox{cos}(\hbar\mR\check{\cal T})
\{
\mbox{Ei}(i\hbar\mR\{\check{\cal T} - {\cal T}\}) - 
\mbox{Ei}(i\hbar\mR\check{\cal T})
\}
\left\}
\left\}
\sqrt{\frac{\mR^2 - \mj^2}{\{2\mR - 1\}\{2\mR + 1\}}}
|\psi_{\sR - 1,\sj}\rangle
\right.
\right.
\right\}
\eeq
Finally, we consider the small ${\cal T}$ regime (small time evolution makes sense for an early-universe application, 
unlike for an atomic application in which one is considering a long-lasting final state;  after that further transitions 
could occur, corresponding to second and higher order terms, which is beyond the scope of this paper).  
Also the motivation for noninteracting initial and final states of atomic physics (involving thus large values of the 
time) is is very much inappropriate in the current analogue-cosmological setting.
\beq
|\psi_{\sR\sj}(I,\Theta,\Phi)\rangle \approx |\psi_{\sR\sj}(\Theta, \Phi) + 
\frac{i}{2} 
\frac{B}{A}
\frac{\{I - I_0\}}{I}
\frac{I_0}{I_{\sc\sh\sa\sr}}
\left\{
\sqrt{\frac{\{\mR + 1\}^2 - \mj^2}{\{2\mR + 1\}\{2\mR + 3\}}}
|\psi_{\sR + 1,\sj}\rangle
+
\sqrt{\frac{\mR^2 - \mj^2}{\{2\mR - 1\}\{2\mR + 1\}}}
|\psi_{\sR - 1,\sj}\rangle
\right\} 
\eeq
for $I_{\sc\sh\sa\sr} := \hbar\sqrt{{-A}/{2}}$ the characteristic quantum moment of inertia scale for this system.  
%
%File R Balance with the Bohr MOI...

This regime is of course computable even in cases for which the full integral cannot be analytically calculated.

%======================================================================================================================
\subsection{Consideration of omitted terms}
%======================================================================================================================

Here we ``justify the consistency" of the previously-made omissions.  
The $h$-equation's neglected pieces are as follows. 
Some of these neglects can be made prior to knowing the explicit form of $|\chi\rangle$, however not all 
can be, since dimensional arguments cannot be used to argue that `averages are small'.  
For 3-stop metroland, 
\beq\hbar^2 \pa^2W \approx O(\hbar)\pa^2W, 
\{\hbar^2/\rho\}\langle \pa_{\rho} \rangle \approx O(\hbar^4)\{B/A\}^2, \mbox{ and }   
\{\hbar^2\}\langle \pa_{\rho}^2 \rangle \approx O(\hbar^4)\{B/A\}^2. 
\eeq 
The $\rho \longrightarrow I$, barred counterpart of the above holds for triangleland. 
For 3-stop metroland, $O(\hbar^2)\langle\pa_{\varphi}^2\rangle$ and the $|\psi_{\sd}^{\st(0)}\rangle$ 
are eigenfunctions, so this one goes as $O(\hbar^2)$ -- an example of an averaged term not being smaller. 
As another example of this, for triangleland, $O(\hbar^2)\langle D_{l}^2 \rangle$, and the $Y$'s are 
eigenfuntions, so this one goes as $O(\hbar^2)$ to leading order. 

\mbox{ }

For 3-stop metroland, $B\langle\mbox{cos}2\varphi \rangle$ in the case that survive the selection rule, goes 
as $O(\hbar) O(B)$.
The triangleland counterpart of this is $B\langle \mbox{cos}\Theta \rangle$, which is a 3-$Y$ integral, 
with $\pm 1$ selection rule, and goes like  $O(\hbar)O(B)$.  
Now, $O(B) > O(\hbar)$ for the time-dependent Schr\"{odinger} equation to make sense, but clearly 
whether this is true or not is down to what regime is selected; these equations have regimes {\sl} 
other than the one in which a time-dependent Schr\"{odinger} equation is a good approximation.

%========================================================================================================
%========================================================================================================
\section{Small but non-negligible backreaction}
%========================================================================================================
%========================================================================================================

%========================================================================================================
\subsection{3-stop metroland case} 
%========================================================================================================

\beq
t_{(1)} - t_{(1)}(0) = \int_{t_{(0)}}^{t_{(0)}} 
\left\{
\frac{1}{2\{\fE - A\rho^2\}}
\int_{0}^{2\pi}\mbox{exp}(i\{\fE_{\sd^{\prime}} - \fE_{\sd}\}{\cal T}/\hbar)
\frac{B\langle \psi_{\sd^{\prime}}|\mbox{cos}2\varphi|\psi_{\sd}\rangle_{\sd}\d t }{4A^2\{1 + 
8|A|^3{\cal T}^2\}^2}
\right\}
= \frac{B}{8A^2}I_{\varphi}^{\st}(\md^{\prime}, \md) J(\beta, {\cal T})  
\eeq
for
\beq
J(\beta, {\cal T}) := \int_0^{{\cal T}_0}
\frac{    \mbox{exp}(i\beta{\cal T})\d{\cal T}   }
     {    \{1 + 8|A|^3{\cal T}^2\}^2\{\fE_{\sH}2|A|\{1 + 8|A|^3{\cal T}^2 - A\}    }
\eeq 
which is evaluable using partial fractions to give integrals in terms of elementary Si and Ci functions 
[and again has a much simpler case for $\beta = 0$].
For small ${\cal T}$, this goes as $-\{B/8A^3\}I_{\varphi}^{\st}(\md^{\prime}, \md)$.

We finally need 
\beq
\langle\chi_{(0)}|\widetilde{\fJ}|\chi_{(0)}\rangle = \frac{B}{4A^2}  \frac{1}{\{1 + 8|A|^2{\cal T}^2\}^2} I_{\varphi}(\md, \md)
\mbox{ } .
\eeq
The $I_{\varphi}$ integral is usually zero, but the d = 1 case survives. 
Then the inhomogeneous term is 
\beq
f(\varphi, {\cal T}) = \frac{B}{A}\frac{1}{4A}
\left\{
\frac{\mbox{cos}2\varphi}{\{1 + 8|A|^3{\cal T}^2\}^2} \pm 
\frac{\hbar^2}{4\{8\fE_{\sH}|A|\{1 + 8|A|^3{\cal T}^2\}  - A\}\{1 + 8|A|^3{\cal T}^2\}^2}
\right\}
\mbox{ } , 
\eeq
the plus case for the cosine solutions and the minus case for the sine solutions.  

\mbox{ } 

The general answer is then the linear combination of the complementary functions for the linear operator in question, 
\beq
L = i\hbar\frac{\pa}{\pa{\cal T}} + \frac{\hbar^2}{2}\frac{\pa^2}{\pa\varphi^2}
\eeq
plus the integral (\ref{Munch}) with the Green's function corresponding to this $L$ and the above $f$ inserted into it.   
This particular example illustrates a backreaction occurring to first order.

%========================================================================================================
\subsection{Triangleland case}
%========================================================================================================

Now, the $t_{(1)}$ integral has a $I_{\Theta\Phi}(\mR, \mR)$ factor which is always zero. 
Thus this example gives no backreaction to first order; I take second order to be beyond the scope of this paper.  
The corrected time-dependent Schr\"{o}dinger equation then has no expectation term, but still does have 
a particular integral from its first term, 
\beq
\widetilde{\fJ}|\psi_0\rangle = \frac{\mbox{cos}\Theta}{\{{\cal T} - \check{\cal T}\}^2}Y_{\sR\sj}(\Theta, \Phi) = 
f({\cal T}, \Theta, \Phi)
\eeq
The general answer is then the linear combination of the complementary functions for the linear operator in question, 
\beq
L = i\hbar\frac{\pa}{\pa{\cal T}} + \frac{\hbar^2}{2}
\left\{
\frac{\pa^2}{\pa\varphi^2} + \frac{1}{\mbox{sin}\Theta}\frac{\pa}{\pa\Theta}
\left\{
\mbox{sin}\Theta
\frac{\pa}{\pa\Theta}
\right\}
\right\}
\eeq
plus the integral (\ref{Munch}) with the Green's function corresponding to this $L$ and the above $f$ inserted into it.

%========================================================================================================
%========================================================================================================
\section{Conclusion}
%========================================================================================================
%========================================================================================================

This paper treats RPM's in parallel to semiclassical quantum cosmology, via the correspondences 
I previously gave in \cite{ScaleQM, 08III}.  
These models illustrate the limitation that semiclassical approximations' validity can be 
configuration space region-dependent (noting that the semiclassical approach's approximations holding 
reduce in the 3-body problem context to the decidedly shaky assumption of the 2-body problem being 
stable within the 3-body problem...)
In particular, we set up 

\mbox{ }

\noindent
1) the negligible-backreaction regime.  
Here, one has a Hamilton--Jacobi equation and then an emergent-time-dependent perturbation of an 
emergent-time-dependent Schr\"{o}dinger equation.

\mbox{ }

\noindent 
2) The small-but-non-negligible backreaction regime. 
Here, one has a Hamilton--Jacobi equation, then an emergent time-dependent Schr\"{o}dinger 
equation, then an expectation-corrected Hamilton--Jacobi equation 
and finally a new emergent-time-dependent Schrodinger equation problem including an inhomogeneous term 
that comes from the lower-order wavefunction.

\mbox{ }

\noindent 
This paper's method of solution for schemes 1) and 2) also builds on my previous suggestion of rectified time \cite{Cones}, 
which I interpret more clearly in the present paper (monotonicity lemma, interpretation as 
the on-shape space version of the time whilst the emergent time itself is the on-relationalspace 
version). 
I solved both schemes 1) and 2) for this paper's specific 3-stop metroland and triangleland RPM toy models, 
albeit I left the solution to 2) in terms of the Green's functions for the emergent-time-dependent Schr\"{o}dinger 
equation on an annulus and a spherical shell, in each case with the time playing the role of radial thickness.

My models provide an example in which the backreaction of the light subsystem on the emergent 
timefunction goes as the strength of the interaction between the heavy and light subsystems, 
and another case in which there fails to be such a linear contribution.  
[This difference is essentially down to the presence of selection rules.] 
In both cases, the $l$-state is altered to first order, albeit in a much simpler way in the latter case 
(no expectation-based interaction term).

As regards the other approximations made, for now I have some qualitative cautions 
(I envisage RPM's as qualitative probes of the Halliwell--Hawking scheme). 

\mbox{ }

\noindent 
I) One has a `dropping higher derivatives debacle': there is a potential danger in ignoring higher derivative terms even 
if they are small (c.f. Navier--Stokes equation  versus  Euler equation in the fluid dynamics). 

\mbox{ }

\noindent 
II) Also, the expectation/averaged terms are often dropped in the literature.  
People in the field often argue that as averaged terms these will be negligibly small. 
However, the present paper's models provide a type of counterexample to this. 
People in the field probably do not want such terms to be around due to 
non-amenability to exact treatment.  
However, in Molecular Physics it is understood that such terms require a {\it self-consistent} 
variational--numerical approach, an example being the iterative techniques of Hartree--Fock.  
Does this Hartree--Fock approach to atomic and molecular physics teach us some things as regards the 
semiclassical approach to Quantum Cosmology?
Some such terms in the light equation can be treated as quite a direct parallel of the original spinless 
Hartree approach, albeit for a time-dependent wavefunction.   
Hartree--Fock is usually like the second equation above in the timeless case, but with expectations in it 
and requiring self-consistent solution rather than merely ignoring the expectation terms.
However, it is sometimes treated in a time-dependent scheme, bringing one closer to the second equation above. 
Some side issues for II) are as follows. 

\mbox{ }

\noindent
i) Is there any promise to additionally incorporating the approximate atom-by-atom product form of 
Hartree--Fock wavefunctions to make a clump-by-clump analysis of inhomogeneous cosmology? 
The nonlinearity of GR is ultimately sure to cause problems here, though this clump-by-clump 
possibility might just be possible within the RPM toy models themselves.

\mbox{ }

\noindent 
ii) Usually in Hartree--Fock theory one has a totally antisymmetric combination of products, but, 
for clumps in a cosmological modelling setting we no longer have motivation to ascribe fermionic statistics.  
There is however a lesser amount of literature on Hartree--Fock with such an omission of fermionic statistics 
to rest upon.  

\mbox{ }

\noindent 
iii) There are also ultimately issues with converting such a method from finite theory to field theory. 
However, Hartree--Fock theory is certainly also familiar in field-theoretic form in Condensed Matter Physics.

\mbox{ }

\noindent One then has the issue of whether the iterative perturbation scheme converges; the nonstandardness of 
the problems at hand may make this hard to check.
There is, however, some possibility of some parallels with inclusion of backreaction due to changes in the state of the nuclei.  

\mbox{ } 

\noindent 
So far, the 2-body in 3-body, Euler in Navier--Stokes and emergent-time-dependent Schr\"{o}dinger equation within 
an emergent-time-dependent Hartree--Fock scheme issues substantiate a qualitative suspicion of some of the approximations 
being made in them.  
As such it may be worth questioning for the moment whether we actually believe in semiclassical quantum 
cosmological calculations.  

\mbox{ } 

\noindent RPM's are also open to checks of the semiclassical approach's assumptions and approximations against 
exact solutions: the system is simple enough that it can be solved without making semiclassical assumptions; 
then in which cases and in which regions does the semiclassical approach give a good approximation to the exact 
solution?
E.g. I calculated this for the 3-stop metroland with HO-like potential in \cite{ScaleQM}.  
Alternatively, a wider range of exactly soluble models including those in \cite{08III} for triangleland, might be 
comparable against the small-time approximation solution 
of their not-exactly-soluble semiclassical schemes 1) and 2).

%==================================================================================================
\subsection{Extensions of this paper to other models} 
%================================================================================================== 

As regards extension to quadrilateralland (which I begin to study in \cite{QShape, QSub} and 
$N$-a-gonland, note that the Hamilton--Jacobi equation for this is the same as that for this 
paper's $3$-stop metroland.  
Thus this paper also gives a start to semiclassical calculations for these models which have further 
uses through their various new nontrivialities.
For, they simultaneously possess nontrivial constraints, a scale variable for the close analogy 
to Quantum Cosmology and nontrivial non-overlapping and hierarchical subsystems that are useful 
for records theoretic and structure formation applications.

\mbox{ }

Midisuperspace (or midisuperspace perturbations about minisuperspace) counterparts of this paper's 
schemes would also be interesting.  
E.g. whether each of multiscalar isotropic GR quantum cosmology, anisotropic GR quantum cosmology and 
perturbatively inhomogeneous GR quantum cosmology admit close analogues of this paper's two schemes, 
whether these are in any such cases soluble, and whether the Hartree--Fock self-consistent approach
extends this far.

%==================================================================================================
\subsection{Other Problem of Time applications of RPM's} 
%================================================================================================== 

\noindent The na\"{\i}ve Schr\"{o}dinger interpretation has been considered in \cite{AF, +tri, 
ScaleQM}, the simple shape space and relational space geometries lending themselves well to the 
characterization of regions over which to evaluate (\ref{NSI}).
RPM's are, furthermore, particularly tractable as regards Records Theory \cite{Records, NOD}. 
This is because 

\mbox{ }

\noindent 1) RPM's have notions of localization in space (e.g. Kendall's notion of 
almost-collinearity \cite{Kendall} and others explained in \cite{AF, +tri, QSub}). 

\mbox{ }

\noindent
2) RPM's have notions of localizaton in configuration space (due to knowing the 
geometry and it being simple; in particular, as explained in the Introduction, 
RPM's have a positive-definite kinetic metric on which suitable such notions of 
localization can be based).    

\mbox{ }

\noindent
3) By QM solvability allowing one to build up a statistical mechanics  
and thus notions of entropy, and negentropy is a reasonable characterization of information, 
RPM's have tractable notions of information, subsystem information, mutual information and so on. 
With these structures in place, one would hope to be able to investigate the extent to which 
records theory can by itself produce a semblance of dynamics or history \cite{EOT}. 

\mbox{ }

\noindent 
RPM's are additionally tractable as examples of Histories Theory \cite{08III, FileR}.
Thus, combining with the present paper's semiclassical approach developments, I have shown that 
one can get far with each of the three strategies 
that I am next proposing to combine.
The next step will involve producing more complicated and genuinely closed-universe versions 
of Halliwell's work on various partial combinations of these three approaches \cite{H99, H03, H09}.  
I outline the form of this next step in more detail in \cite{ASharp}.  

\mbox{ }

Finally, some sorts of `tempus ante quantum' approaches to the Problem of Time are also 
tractable for RPM's \cite{06II, SemiclI, FileR}  (albeit it runs into 
difficulties at the quantum level).  
In particular, one has an `Euler' or dilational time $t^{\sE\su\sll\se\sr} = \sum_e{\bf R}^e\cdot{\bf P}_e$ 
\cite{06II, SemiclI} in close analogy with the York Time of GR, $t^{\sY\so\sr\sk} = \frac{2}{3}h_{ij}\pi^{ij}/\sqrt{h}$.  
For the present paper's models I consider this for comparison in \cite{FileR}.  

\mbox{ } 

\noindent{\bf Acknowledgements} I thank my wife Claire, 
Amelia, Sophie, Sophie, Anya, Emily, Tea and Alicia.   
Professor Jonathan Halliwell for correspondence.  
Professor Enrique Alvarez and Dr Julian Barbour for reading earlier drafts of this article.  
Professors Enrique Alvarez, Jeremy Butterfield, Belen Gavela, Marc Lachi\`{e}ze-Rey, Malcolm MacCallum and 
Don Page for help with my career.

%=====================================================BIBLIOGRAPHY==========================================================================

\end{document}